\documentclass[12pt,preprint]{aastex62}
\usepackage{CJK}
\usepackage{xcolor}
\usepackage{multirow}
\begin{document}
\begin{CJK*}{UTF8}{gbsn}
\title{Followup ground-based observations of the dwarf nova KZ\,Gem}
\author[0000-0002-4280-6630]{Zhibin Dai (戴智斌)}
\affiliation{Yunnan Observatories, Chinese Academy of Sciences, 396 Yangfangwang, Guandu District, Kunming, 650216, China.}
\affiliation{Key Laboratory for the Structure and Evolution of Celestial Objects, Chinese Academy of Sciences, 396 Yangfangwang, Guandu District, Kunming, 650216, China.}
\affiliation{Center for Astronomical Mega-Science, Chinese Academy of Sciences, 20A Datun Road, Chaoyang District, Beijing, 100012, China.}
\affiliation{University of Chinese Academy of Sciences, No.19(A) Yuquan Road, Shijingshan District, Beijing, 100049, China}
\affiliation{Key Laboratory of Optical Astronomy, National Astronomical Observatories, Chinese Academy of Sciences, Beijing 100101, China}
\author[0000-0003-4373-7777]{Paula Szkody}
\affiliation{Department of Astronomy, University of Washington, Seattle, WA, 98195, USA.}
\author{John R. Thorstensen}
\affiliation{Department of Physics and Astronomy, Dartmouth College, 6127 Wilder Laboratory Hanover, NH 03755, USA.}
\author{N. Indika Medagangoda}
\affiliation{Astronomy Division, Arthur C. Clarke Institute for Modern Technologies, Sri Lanka.}
\correspondingauthor{Zhibin Dai}
\email{zhibin\_dai@ynao.ac.cn}

\begin{abstract}
We present spectroscopy of stars in the immediate vicinity of the dwarf nova (DN) KZ Gem to confirm its identification, which had been ambiguous in the literature. Analysis of 73 radial velocities spanning from 2014 to 2019 provides a high-precision orbital period of 0.2224628(2)\,d ($\sim5.34$\,hr) and shows KZ\,Gem to be a double-lined DN. Time series photometry taken from 2016 to 2018 shows a variable double-hump modulation with a full amplitude of $\sim0.3$\,mag, along with five Gaussian-like transient events lasting $\sim30$\,min or more. Using the light curve code XRBinary and nonlinear fitting code NMfit, we obtain an optimized binary model of the dwarf nova (DN) KZ Gem, from time series photometry, consisting of a Roche-lobe-filling K type dwarf with a mass transfer rate of $2.7\,-\,7.9\times10^{-10}\,{\rm M}_{\odot}\,{\rm yr}^{-1}$ to a large, cool and thick disk surrounding a white dwarf, in an orbit with an inclination of $51^{\circ}.6(\pm1^{\circ}.4)$. Two hotspots on the disk are demonstrated to cause the observed variations in the ellipsoidal modulations from the secondary star. This physical model is compatible with the Gaia distance of KZ\,Gem.
\end{abstract}

\section{Introduction}

\citet{hof66} and \citet{kuk68} first discovered KZ\,Gem as a variable star a half century ago. It is listed in various catalogs of cataclysmic variables (CVs; \citealt[e.g.][]{dow93,dow97,dow01,rk724}) as a dwarf nova (DN). Since KZ\,Gem falls within the field of view of the K2 Campaign 0 (K2-C0), it was proposed as a CV target and observed in long cadence (30\,min sampling) mode. In the K2 variable catalog\footnote{\url{http://archive.stsci.edu/k2/hlsp/k2varcat/search.php}}, KZ\,Gem was listed as OTHPER (i.e., other periodic and quasi-periodic variables \citep{arm15,arm16}). The ``self-flat-fielding" (SFF) corrected light curve of KZ\,Gem \citep{van14a,van14b} shows an ellipsoidal-like effect. \cite{dai17} applied a phase-correction method to these data to show that the orbital period is 0.22242\,d, almost exactly twice the orbital period of 0.11122\,d listed in RKcat (Edition 7.24; \citealt{rk724}). A DN outburst of KZ\,Gem was detected in 2015 January by \citet{lan16}, but without a published spectrum, KZ\,Gem remained a poorly understood DN.

\begin{figure}
\centering
\includegraphics[width=14.0cm]{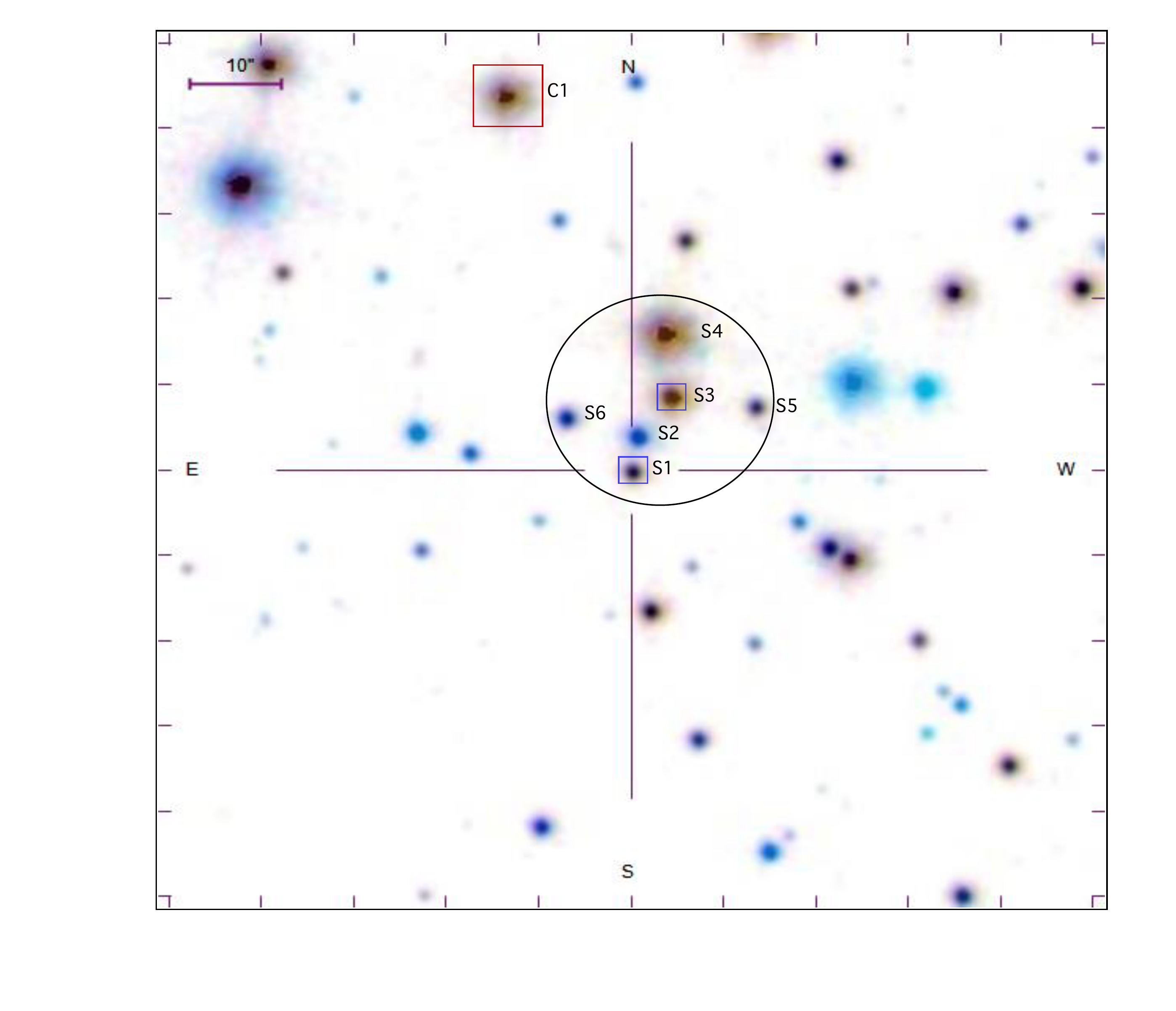}
\caption{\small{Finding chart of KZ\,Gem from the SDSS DR12 \citep{ala15}. The black circle is $15''$ in radius. The star marked S1 corresponds to the coordinates in SDSS, SIMBAD, and K2, and S3 corresponds to the coordinates in RKcat, VSX, and Gaia. We show here that S3 is the dwarf nova. The comparison star C1 used for the XLO and APO photometry is marked with a red square; the MDM photometry used S1.}}
\label{findingchart}
\end{figure}

The coordinates for KZ\,Gem given in RKcat and the VSX (Variable Star Index) databases of the AAVSO\footnote{\url{https://www.aavso.org/vsx/index.php?view=detail.top&oid=14553}} differ by $9''$ from those in SDSS, SIMBAD, and K2. Because the field of KZ\,Gem is crowded, a difference of $9''$ is large enough to cause an incorrect identification. Fig.~\ref{findingchart} shows 6 stars within $15''$ of KZ\,Gem; Table 1 lists the SDSS coordinates of these stars. The two different coordinates listed for KZ\,Gem correspond to stars S1 and S3, which are marked by blue rectangles in Fig.~\ref{findingchart}. The crowding in the field also leaves open the possibility that the ellipsoidal modulation found by \citet{dai17} in the K2 light curve might arise from a neighboring star rather than KZ\,Gem.     

This paper presents ground-based spectroscopy of four of the nearby stars (S1, S2, S3, and S4) and photometry of S3. In Section 2, we confirm that the identification of S3 with KZ\,Gem, and in Section 3 we obtain a high-precision orbital period from radial velocities. Folding with this period, the light-curve morphology, models and five transient events are discussed.

\section{Observations}

Table 2 gives a journal of the observations.

\subsection{Spectroscopy}
\label{spectroscopysection}

Our spectra are from four different instruments on three different telescopes. (1) With the Double Imaging Spectrograph (DIS) on the Apache Point Observatory (APO) 3.5m telescope, we obtained spectra on 5 nights using the blue and red channels simultaneously. Gratings B1200/R1200 gave a dispersion of $\sim0.6$\,\AA\ ${\rm pixel}^{-1}$. (2) With the Beijing Faint Object Spectrograph (BFOSC) and grating G4, on the Xinglong Observatory (XLO) 2.16m telescope, we obtained spectra on 3 nights with a resolution of $\sim2000$ (2.97\,\AA\,${\rm pixel}^{-1}$). (3) With the modspec on the Hiltner 2.4m telescope at Michigan-Dartmouth-MIT (MDM) Observatory in Kitt Peak, Arizona, we obtained spectra on 12 nights spread over four observing runs (2014 December, 2016 January, 2016 February, and 2018 November). A 600 line\,mm$^{-1}$ grating gave $\sim3.5$\AA\ resolution from 4310 to 7500 \AA, with vignetting toward the ends of the range. (4) Also with the 2.4m MDM telescope, we used the Ohio State Multi-Object Spectrograph (OSMOS) on 6 more nights in 2018 December and 2019 January, with the aim of disambiguating the long-term velocity cycle count. The spectra covered from 3970 to 6870 \AA\ with a resolution of 3 \AA\ FWHM. Flux standards were observed when appropriate, and comparison lamps were observed to maintain accurate wavelength calibration. All spectra were reduced using IRAF\footnote{IRAF is distributed by the National Optical Astronomy Observatory, which is operated by the Association of Universities for Research in Astronomy (AURA) under cooperative agreement with the National Science Foundation.}.

As we show below, the spectrum of the star we identify with KZ Gem shows a contribution from a late-type secondary star. We measured absorption velocities in the MDM spectra by cross-correlating the spectra against a composite template spectrum, which was originally composed by taking 76 spectra of late-type velocity standard stars, shifting them to zero velocity, and averaging. The cross-correlation was either from the rvsao package \citep{kurtzmink}, which implements the algorithm developed by \citet{ton79}, or the fxcor task in the IRAF rv package.

To estimate the secondary star's spectral type and contribution, we shifted the individual MDM 2.4m exposures to the secondary's rest frame and averaged the shifted spectra.  We have a collection of archival spectra of K- and M-type main sequence stars classified by \citet{boe76} and \citet{kee89}, taken with the same instrument setup as our KZ Gem spectra, and also shifted to zero radial velocity.  We estimated the secondary's spectral type and its fractional contribution by scaling the main sequence spectra and subtracting them from the KZ Gem spectrum, interactively trying different spectral types and scaling factors until the secondary-star absorption lines were cancelled as well as possible.

\subsection{Identification of KZ\,Gem} 

\begin{figure}
\centering
\includegraphics[width=14.0cm]{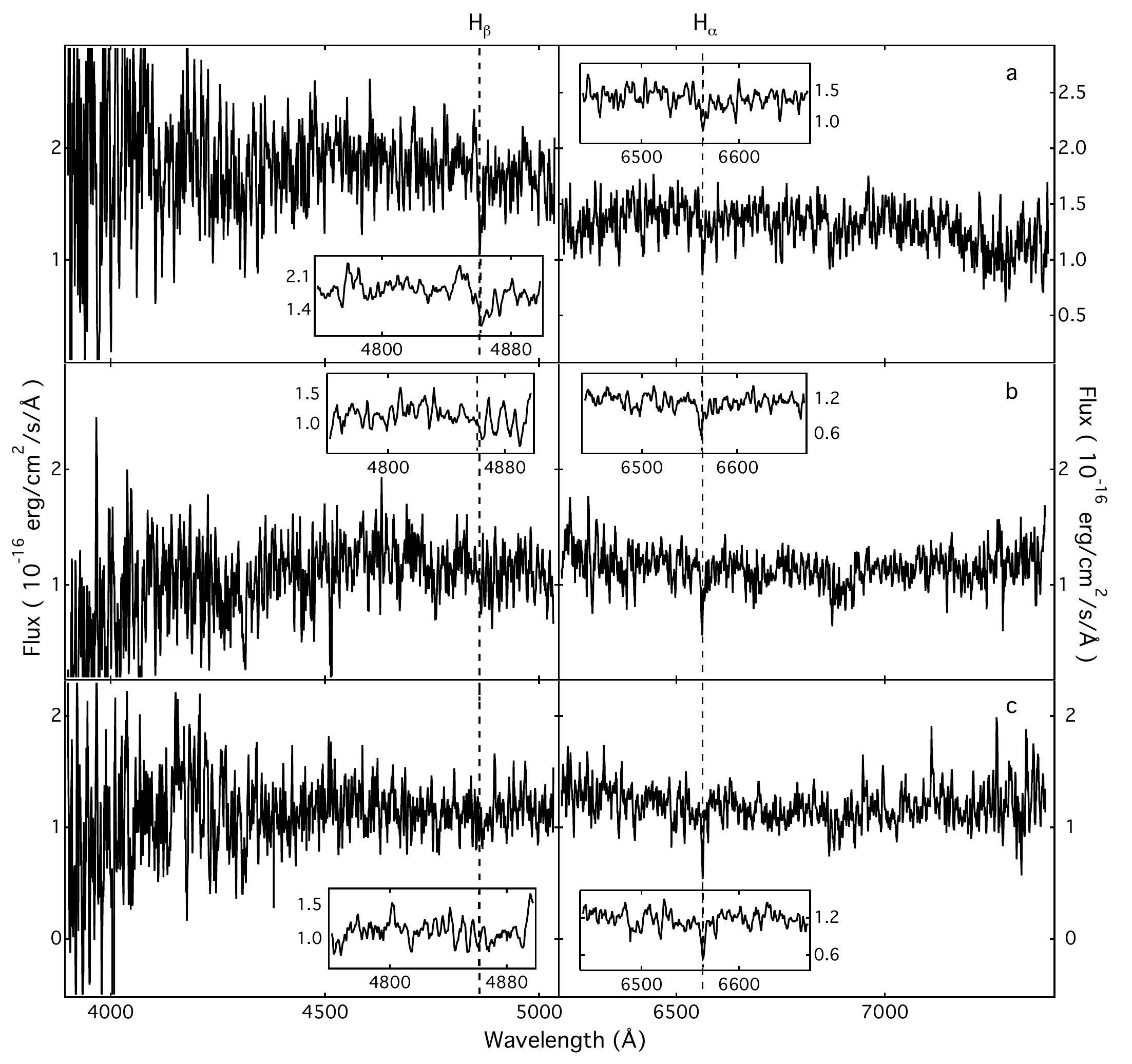}
\caption{\small{Blue and red spectra of S1 from the APO 3.5m telescope are shown in the left and right panels, respectively. From top to bottom, the spectra are from 2016 Dec. 05 (panel a) and 2018 Jan. 15 (panels b and c). The H$\alpha$ and H$\beta$ absorption lines are smoothed and marked by vertical dashed lines. The insets show the zoomed-in lines. H$\beta$ is not visible in the middle and bottom blue spectra.}}
\label{apostar1}
\end{figure}

\begin{figure}
\centering
\includegraphics[width=14.0cm]{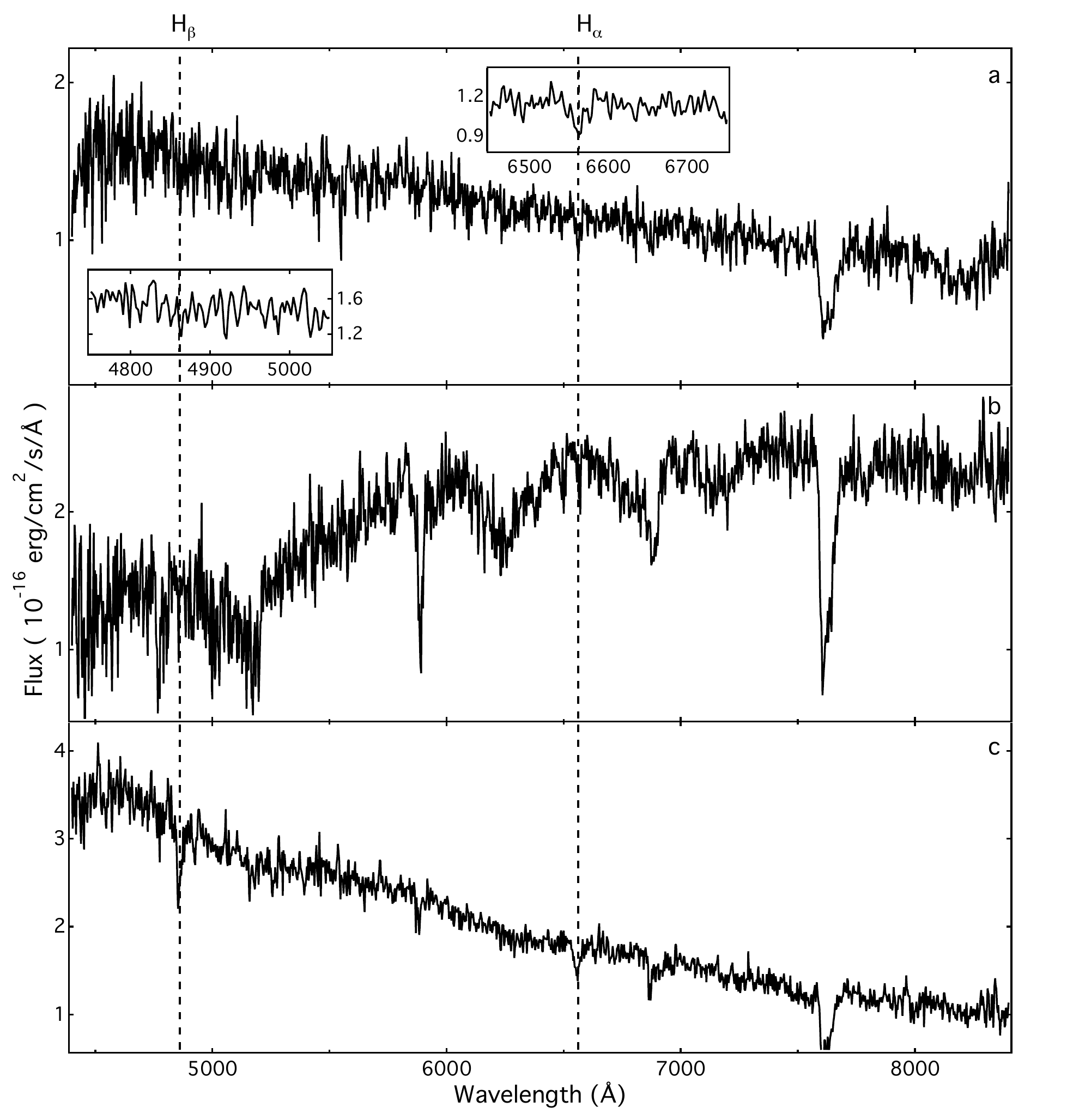}
\caption{\small{The spectra of S1, S2 and S4 located along the north-south direction were simultaneously taken by XLO 2.16m telescope on 2017 Jan. 22. In the top panel, the H$\alpha$ and H$\beta$ absorption lines are smoothed and marked by vertical dashed lines. The zoomed-in lines are shown in the insets. The H$\beta$ absorption line is not detected, as in the blue APO spectra shown in panels b and c of Fig.~\ref{apostar1}.}}
\label{xlostars124}
\end{figure}

Inspection of Fig.~\ref{findingchart} indicates that spectra of stars S1, S2 and S4 can be taken at the same time by using a wide slit along the north-south direction. This was accomplished on 2017 January 22. Since S1 is faint, all spectra of S1 were smoothed by a running boxcar of 5 and 3 pixels for the APO and XLO spectra, respectively. Despite the low S/N, both APO and XLO spectra of S1 (shown in the three panels of Fig.~\ref{apostar1} and the panel a of Fig.~\ref{xlostars124}) illustrate similar features of a sloping linear continuum from the blue to the red with shallow and marginal H$\alpha$ and H$\beta$ absorption lines and a lack of emission lines. Note that the H$\beta$ absorption line shown in the blue APO spectrum taken on 2016 December 05 is undetected in the following two APO spectra and the XLO spectrum. The spectra of S1 are atypical for a quiescent DN. The simultaneous spectra of the other two targets, S2 and S4 (shown in the two panels b and c of Fig.~\ref{xlostars124}, respectively), also only show absorption rather than emission lines; S2 appears to be an early M-type star, and S4 is consistent with a G-type star. There is, therefore, no indication that S1, S2, and S4 are dwarf novae.

\begin{figure}
\centering
\includegraphics[width=14.0cm]{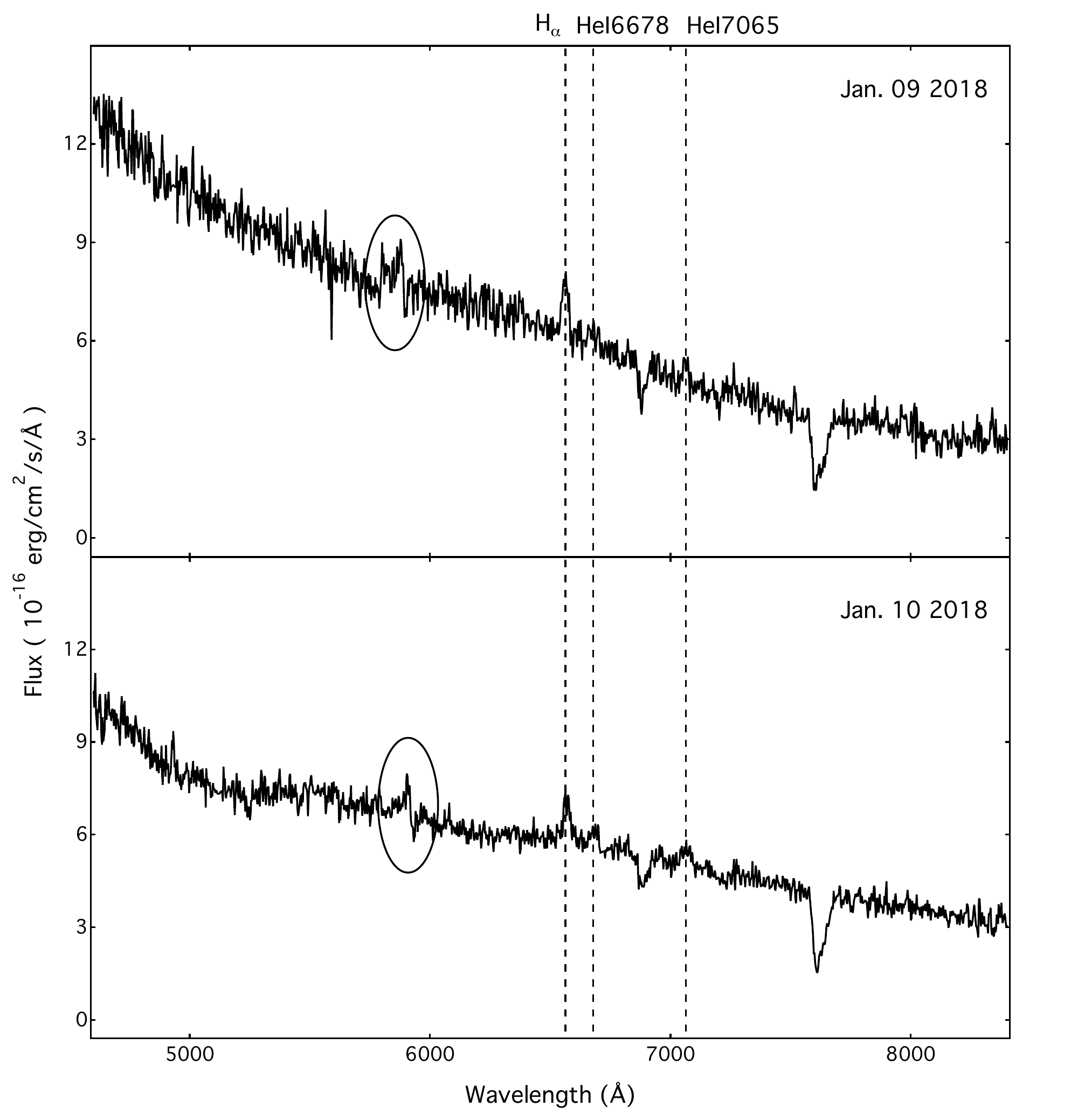}
\caption{\small{Two XLO spectra of S3 taken on 2018 Jan. 09 and 10 are shown in the upper and lower panels, respectively. The blended lines of the \ion{He}{1} $\lambda$ 5876 emission line and the weak absorption line NaD $\lambda$ 5893 are marked by the black ovals.}}
\label{xlostar3}
\end{figure}

\begin{figure}
\centering
\includegraphics[width=14.0cm]{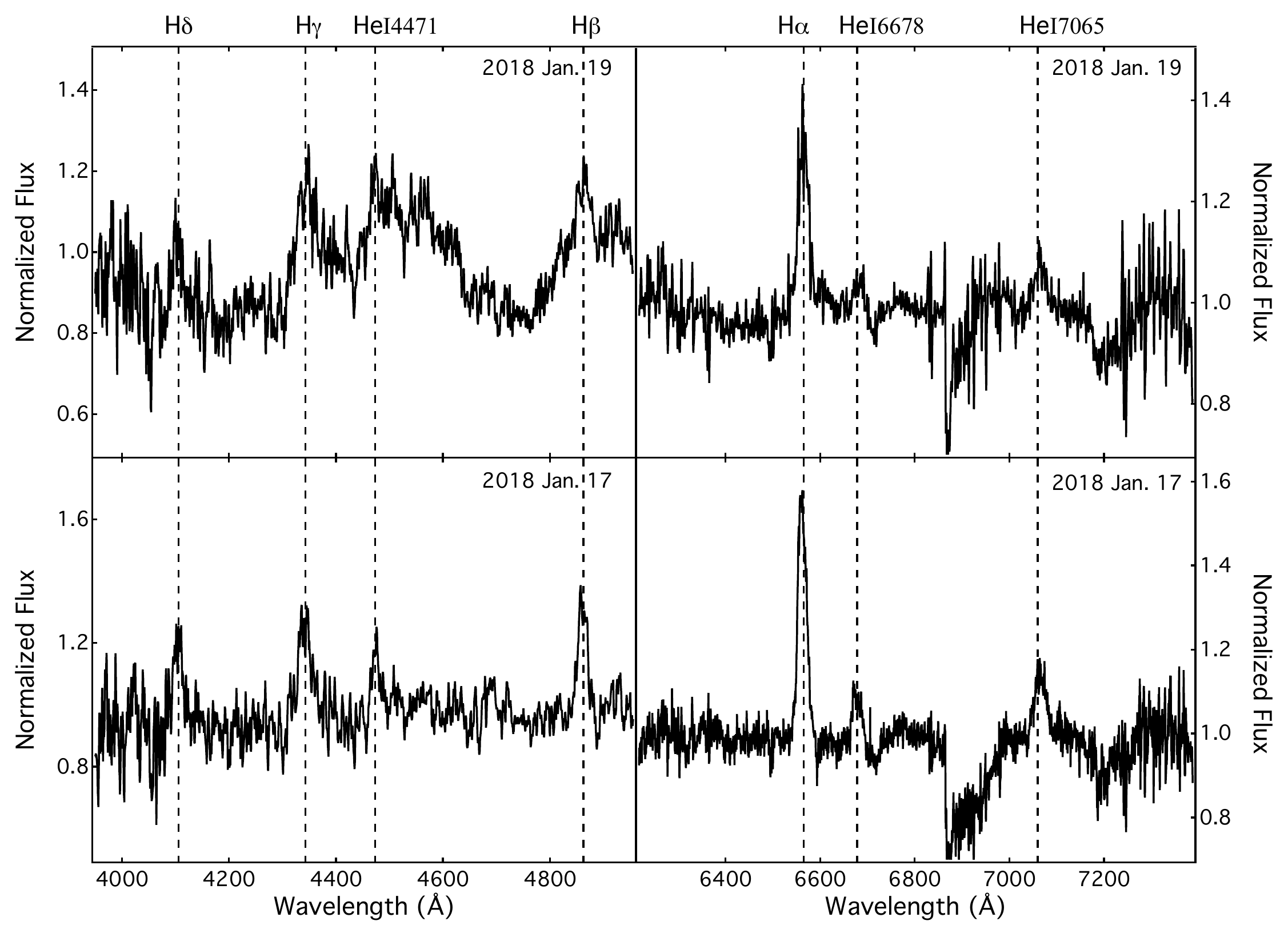}
\caption{\small{From top to bottom, mean APO spectra of S3 taken on 2018 Jan. 17 and 19 are shown. Blue and red spectra are plotted in the left and right panels, respectively.}}
\label{apostar3}
\end{figure}

\begin{figure}
\centering
\includegraphics[width=19.0cm]{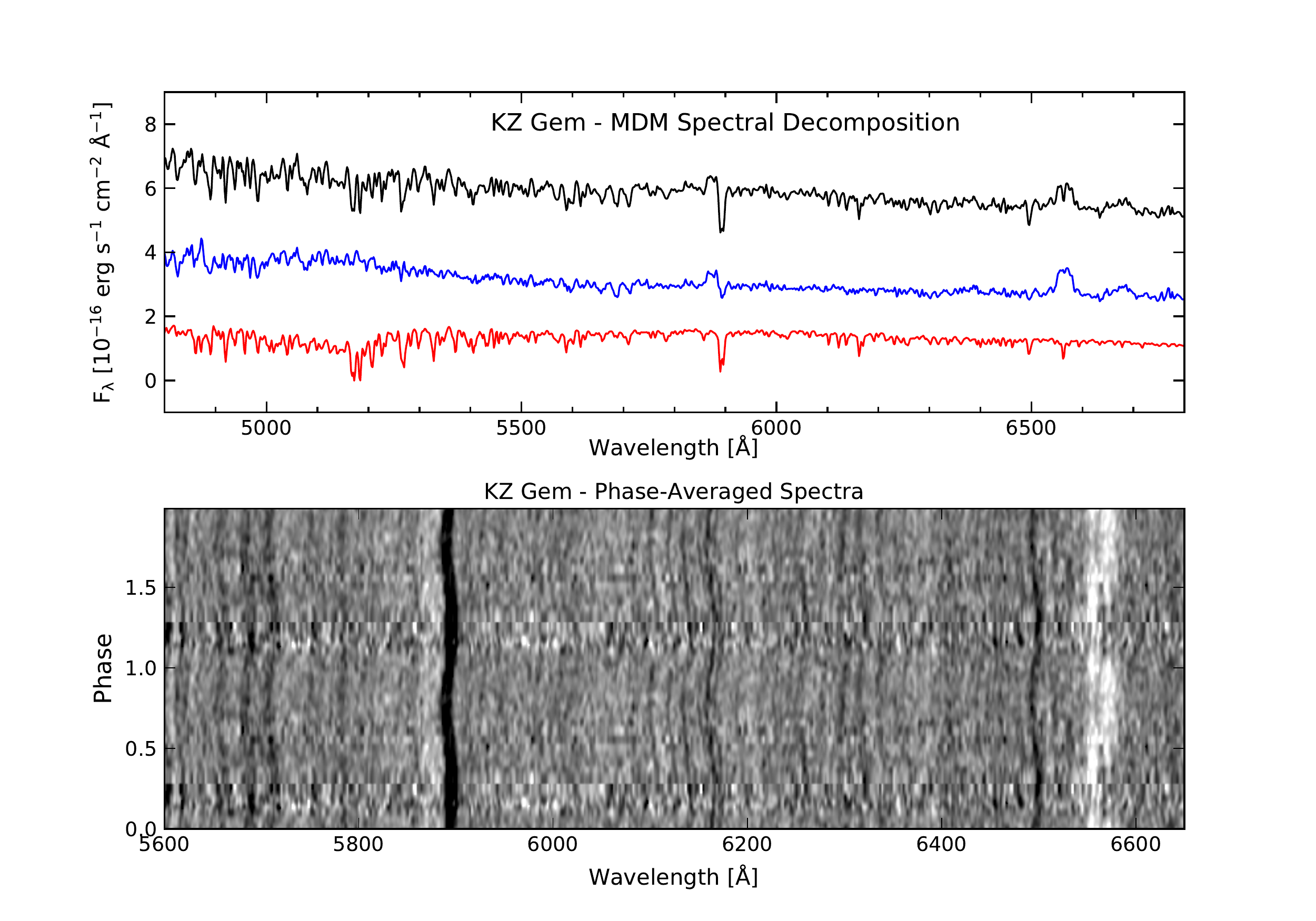}
\caption{\small{Upper panel: (Top trace; black) Mean MDM spectrum of KZ\,Gem in the rest frame of the secondary star.  (Middle trace; blue) Mean spectrum after a scaled spectrum of a K2V star has been subtracted. (Lower trace; red) The scaled K2V star that was subtracted, shifted downward by 1.5 units to avoid overlap with the subtracted spectrum. Lower panel: rectified spectra from the MDM modspec arranged to form a phase-averaged greyscale image. Black and white colors are set to 0.8 and 1.2 times the continuum, respectively. The degrading signal-to-noise near phase 1.2, and a sharp discontinuity around phase 1.3 are artifacts of uneven phase coverage.}}
\label{mdmspecs}
\end{figure}

In contrast, the two XLO spectra of S3 taken on January 09 and 10, 2018 (Fig.~\ref{xlostar3}) clearly show broad H$\alpha$ emission superimposed on the continuum, along with a higher blue flux level. \ion{He}{1} emission lines are marginally visible at $\lambda\lambda$ 5876, 6678, and 7065. These features imply that S3 is the DN, confirming the coordinates listed in RKcat, VSX and Gaia\footnote{The Gaia DR2 gives $\alpha\,=\,06^{\rm h}\,53^{\rm m}\,02.^{\rm s}457, \delta\,=\,+16^{\circ}\,39'\,58''.67$ for this star \citep{are18}.}. In both spectra, the \ion{He}{1}\,$\lambda$ 5876 emission line is blended with weak NaD absorption (as it is in DN SDSS\,J063213.1+253623, or J0632+2536; \citealt{dai16}), and the \ion{He}{1} $\lambda$ 5876 emission observed on 2018 January 09 appears doubled. The averaged blue and red high-resolution APO spectra are shown in Fig.~\ref{apostar3}. H$\alpha$, H$\beta$ and H$\gamma$ are visible in emission in the spectra taken on 2018 January 17 and the lower two panels of Fig.~\ref{apostar3} clearly show neutral He emission at $\lambda\lambda$ 4471, 6678, and 7065. Despite the low S/N of the spectra taken on 2018 January 19 (due to weather and bad seeing), H$\alpha$ is visible in the bottom right panel of Fig.~\ref{apostar3}. Table 3 lists the emission equivalent widths of H$\alpha$ and H$\beta$. The upper panel of Fig.~\ref{mdmspecs} shows the mean of MDM spectrum of S3, which appears similar to the two XLO spectra. All the XLO, APO and MDM spectra have similar continuum flux and moderate Balmer emission lines, indicating KZ\,Gem was in a similar accretion state during the observations.

\subsection{Photometry}

Our differential time-series photometry is from the 1\,K FlareCam with $1''.3\,{\rm pixel}^{-1}$ when binned $2\times2$ mounted on the APO 0.5m Astrophysical Research Consortium Small Aperture Telescope (ARCSAT), and Andor CCD cameras on the XLO 0.85m and MDM McGraw-Hill 1.3m telescopes. We obtained 14 light curves spanning almost three years. For the XLO and ARCSAT data, we used star C1 (Fig.~\ref{findingchart}) as the comparison star, while for the MDM data we used S4, which was well-resolved from the DN. The MDM 1.3m data from 2016 February 13, 14, and 15 were obtained with a GG420 filter\footnote{The GG420 filter blocks light at wavelengths $<$4200\,\AA\ and is approximately equivalent to no filter.} (hereafter, GG420-band), and no filter was used in four XLO observations with the 0.85m and 2.16m telescopes. The XLO and ARCSAT data were obtained using the Point Source Function (PSF) in IRAF standard routines due to the contamination from the two nearby objects S2 and S4. Five light curves were also obtained with Johnson-Cousins V filters (hereafter, V-band) on the APO 0.5m and MDM 1.3m telescopes. The ARCSAT data taken on 2016 November 29 was the only light curve obtained in a sdss $g$ filter.

\section{Results and Discussion}

\subsection{Period analysis}

Since the K2 data are defocused and KZ\,Gem is located in a crowded field, the wide aperture shown in the K2 image plot of KZ\,Gem in the Mikulski Archive for Space Telescopes (MAST) \footnote{\url{https://archive.stsci.edu/prepds/k2sff/html/c00/ep202061320.html}} \citep{van14b,van15} indicates that the SFF corrected light curve of KZ\,Gem includes the flux from several nearby stars. Hence, the orbital period of 0.22242\,d ($\sim5.34$\,hr) derived by \citet{dai17} from the blended K2 data needed further verification.

\begin{figure}
\centering
\includegraphics[width=14.0cm]{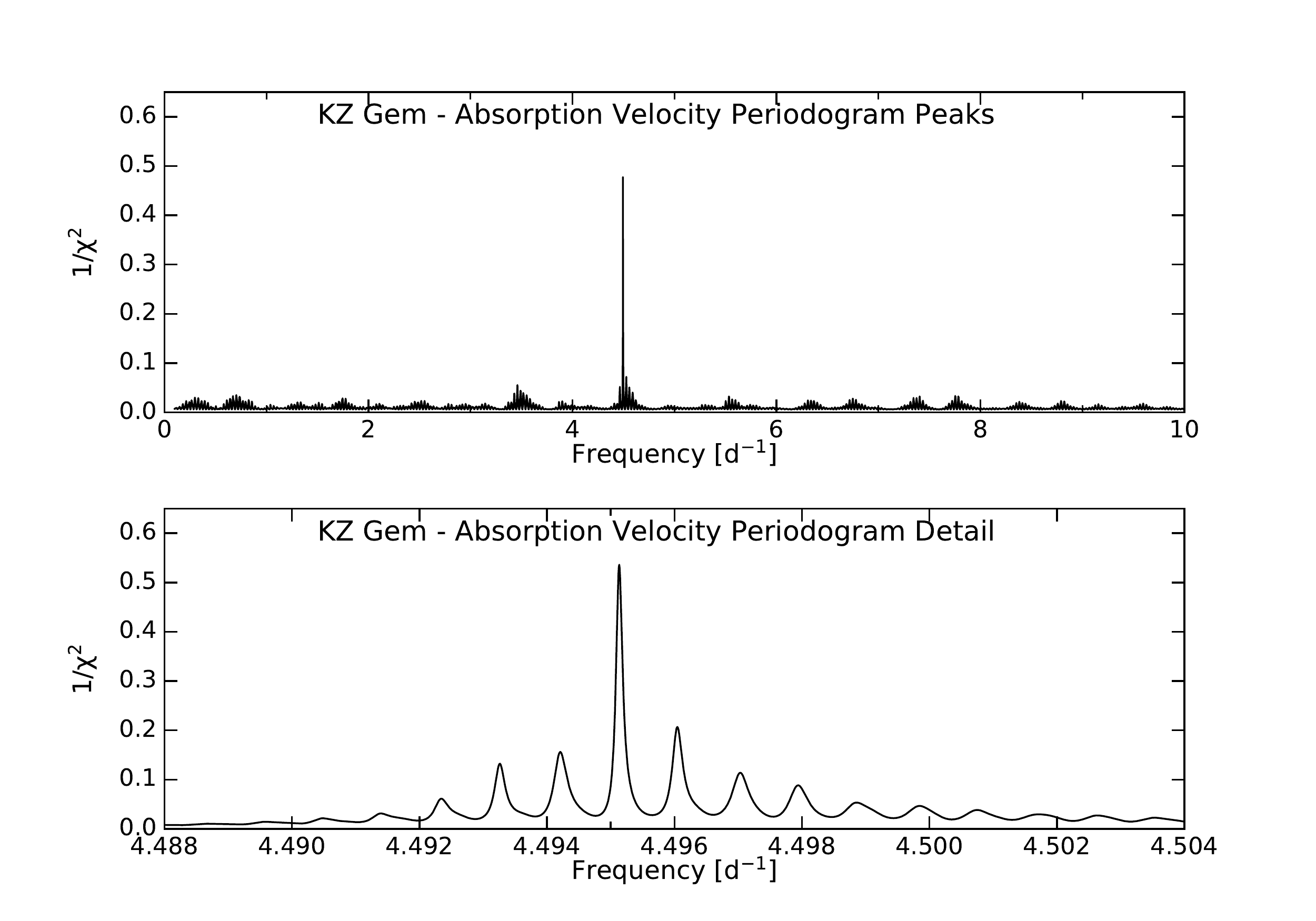}
\caption{\small{Periodogram of the absorption line velocities for KZ Gem. The top panel shows local maxima of the function joined by straight lines, to compress the plotting of the extremely fine grid of test frequencies necessitated by the $\sim$\,3-year span of the data. The lower panel shows the full periodogram, highly oversampled, in the neighborhood of the obvious peak. The flanking peaks correspond to frequencies differing by one cycle per $\sim$\,1090\,d.}}
\label{pgrm}
\end{figure}

\begin{figure}
\centering
\includegraphics[width=14.0cm]{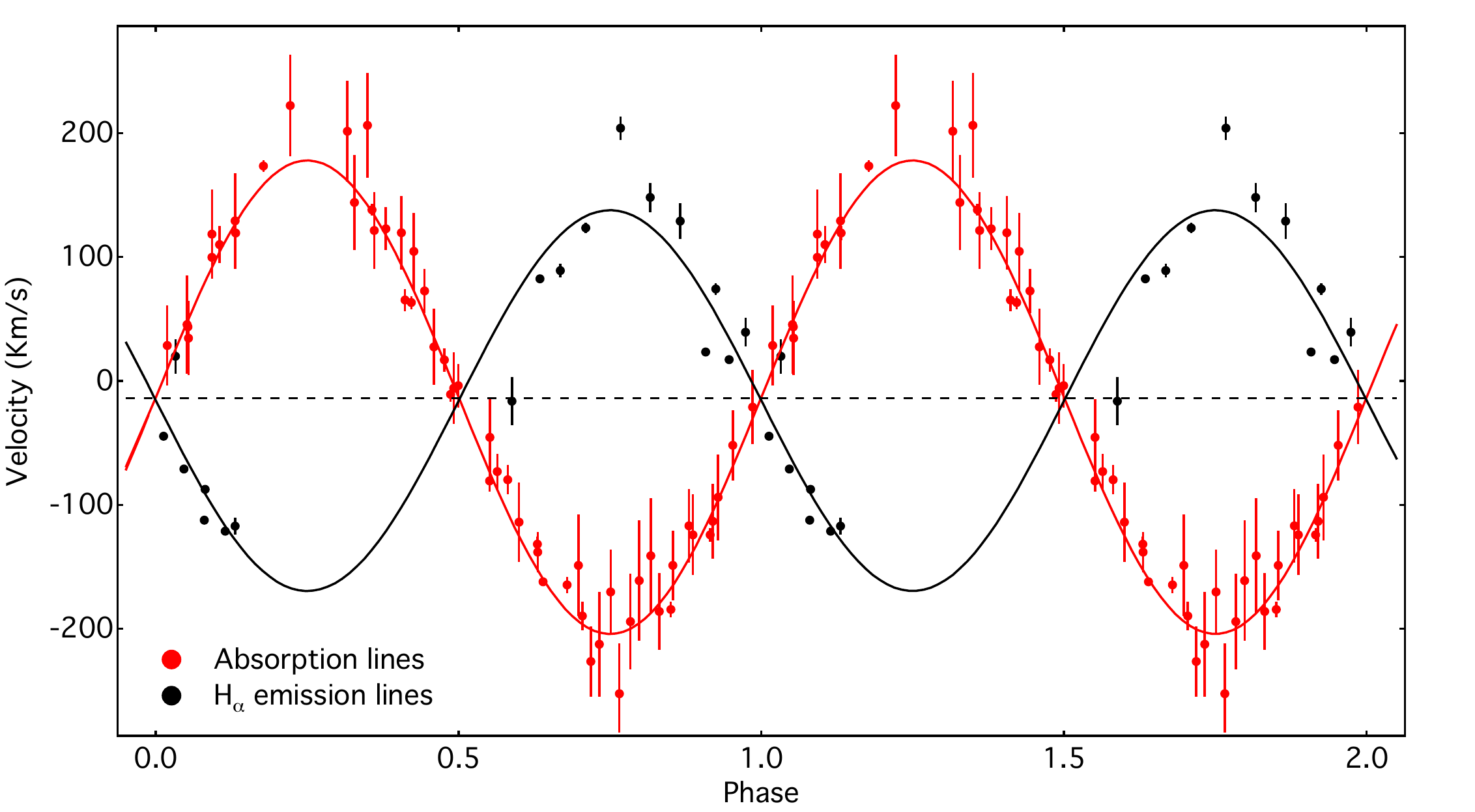}
\caption{\small{Radial velocities and best-fitting sinusoids folded using the ephemeris in Equation 1. The black circles and curve are from the APO emission (H$\alpha$) velocities, and the red are from MDM cross-correlation (absorption) velocities.}}
\label{foldedvels}
\end{figure}

We were able to measure cross-correlation velocities of the secondary star in 55 of our MDM 2.4m spectra. The APO spectra did not yield cross-correlation velocities. We excluded the NaD $\lambda$ 5893 blended absorption feature from the correlation region due to possible confusion with nearby \ion{He}{1} $\lambda$ 5876 emission. To search for periods, we constructed least-squares sinusoidal fits on a sufficiently dense grid of trial frequencies over the range of typical CV periods. Only a single period, corresponding to a single choice of cycle count between the observing runs, gave an acceptable result. A sinusoidal fit to the velocities of the form
\begin{equation}
Vel(t)\,=\,\gamma\,+\,K\sin[2\pi(t-T_{0})/P_{\rm orb}]
\end{equation}
yielded the parameters listed in Table 4. $T_{0}$ is the time when the secondary star passes from blue to red through the mean velocity (i.e., the inferior conjunction of the secondary star). The orbital period found in the search, and refined using Equation 1, is $P_{\rm orb}=0.2224628(2)$\,d, near 5.34\,hr, in good agreement with the period derived from the K2 photometry. The uncertainty is small because the cycle count is unambiguous over more than 4 years. To further verify this derived period, we carried out an absorption velocity periodogram shown in Fig.~\ref{pgrm} using a `residual-gram' method described by \citet{tpst}. A significant peak of this periodogram is at the frequency of 4.4951\,d$^{-1}$ coincident with $P_{\rm orb}$. Although the APO spectra could not contribute to the absorption line fit, the 18 APO H$\alpha$ emission line velocities, obtained over a $\sim46$\,hr time base, corroborate the result, giving $P_{\rm orb}=0.225(3)$\,d, consistent with the absorption-line result. Fixing the period of the emission-line fit to that derived from the absorption lines produces a scatter ($\sigma\sim44\,{\rm km}\ {\rm s}^{-1}$), which is larger than typically found. The best-fitting emission velocity curves shown in Fig.~\ref{foldedvels} display large deviations around phase $0.75-1.0$, when a hotspot would have a maximum contribution. Hence, this may be caused by a hotspot. The ephemeris for the inferior conjunction of the normal star, $T_{\rm inf}$, derived from the absorption line velocities is 
\begin{equation}
T_{\rm inf}\,=\,\hbox{BJD}\,2457434.9280(7)\,+\,0^{\rm d}.2224628(2)\,E,
\end{equation}
where E is the cycle number, and the time base is UTC.

Using this ephemeris, we phased and stacked the 2019 January and February MDM modspec data to create a phase-resolved greyscale image (lower panel of Fig.~\ref{mdmspecs}). Before averaging, we rectified the spectra and edited out cosmic rays and other artifacts. The many absorption features of the late-type secondary are seen Doppler-shifting back and forth with phase. The H$\alpha$ emission line moves in anti-phase to the absorption lines, consistent with an origin in the disk/hotspot surrounding a white dwarf. Assuming that the motion of the emission lines traces the motion of the white dwarf like a long-period DN TT\,Crt \citep{szk92}, the mass ratio of the two component stars may be roughly estimated to be 0.81 using the relationship,
\begin{equation}
q\,=\,\frac{M_{\rm rd}}{M_{\rm wd}}\,\sim\,\frac{K_{\rm em}}{K_{\rm abs}},
\end{equation}
where $K_{\rm em}$ and $K_{\rm abs}$ correspond to the amplitudes of the best-fitting radial velocity curves derived from the emission and absorption lines. A range of $q=0.74-0.88$ is derived by considering the $2\sigma$ errors of $K_{\rm em}$ and $K_{\rm abs}$ listed in Table 4, though larger systematic errors are possible.

\subsection{Secondary Spectrum}

We used the parameters in Table 4 to shift the MDM 2.4m spectra to the rest frame of the secondary as described in Section \ref{spectroscopysection}.  The \citet{gre18} 3-dimensional reddening maps show little extinction in this direction out to $\sim2$\,kpc, so we did not apply any reddening correction to the spectrum before decomposing. The decomposition process yielded acceptable results for spectral types K0 through K5, and suggest that around half the light in the 5000-6500\,\AA\ region arises from the secondary star.  Fig.~\ref{mdmspecs} shows one of the most successful of the decomposition, for a K2V star HD109011, scaled to a flux equivalent to $V=17.9$.  The flux calibration is typically accurate to $\sim0.2$\,mag, limited by cloud and losses at the $1''.1$ spectrograph slit.

\subsection{Light-curve morphology}

\subsubsection{The variable orbital modulation}

Compared with the high-precision orbital period derived from the velocities, a periodogram derived from our differential time-series photometry, based on the Lomb-Scargle periodogram method \citep{lom76,sca82}, shows a notable period of 0.484\,d almost twice $P_{\rm orb}$. Although a trivial peak at the period of 0.221\,d can also be found in this periodogram, a large discrepancy of $\sim2$\,min from $P_{\rm orb}$ implies that the orbital modulation of KZ\,Gem is complex and noisy.

\begin{figure}
\centering
\includegraphics[width=14.0cm]{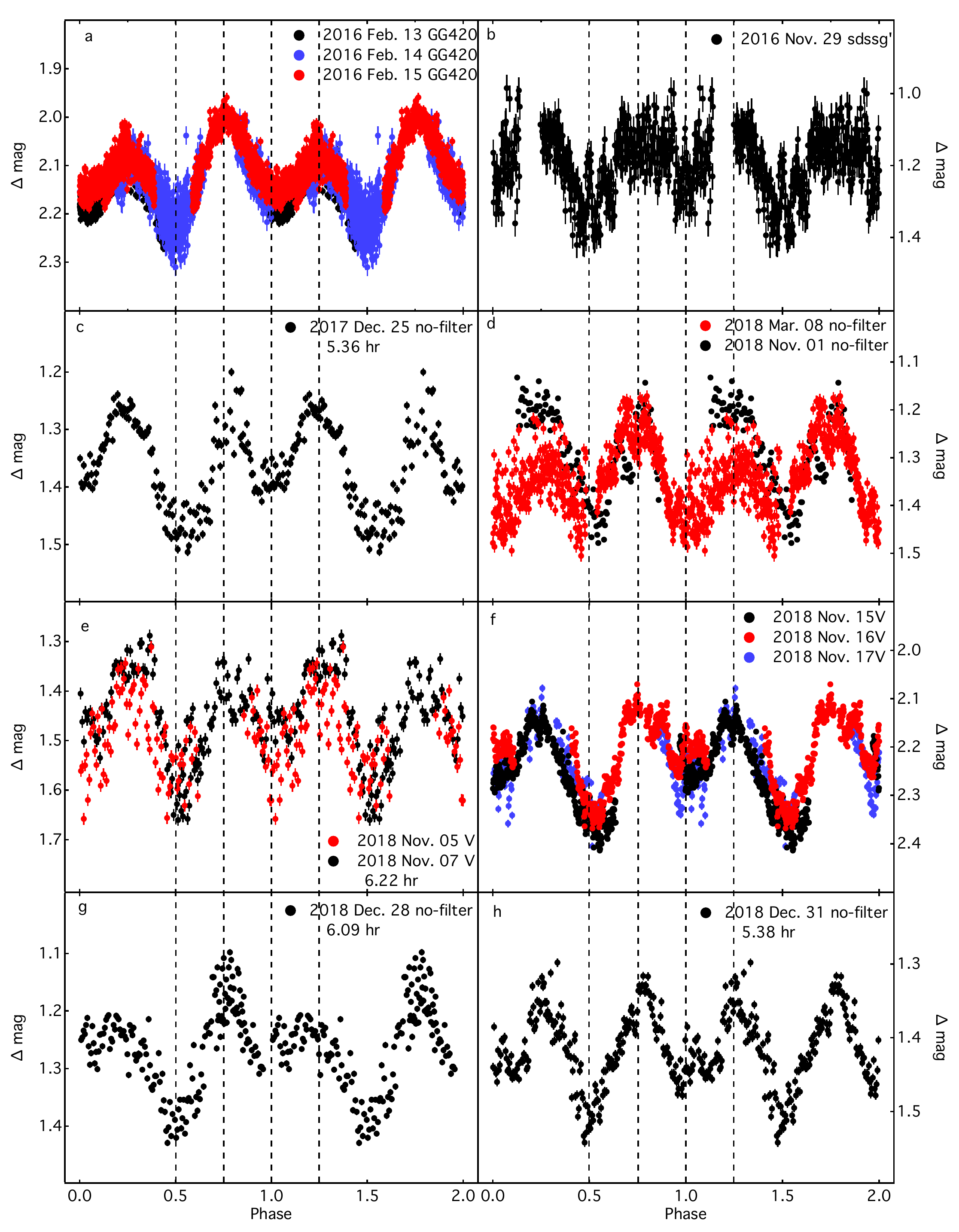}
\caption{\small{Fourteen differential light curves in sdss $g$, GG420, no-filter and V bands, phased with the ephemeris in Equation 2. The durations of light curves longer than the orbital period are listed with the legends. The error bars indicate the standard deviation (STD) of the magnitude difference between the comparison and check stars.}}
\label{lightcurves}
\end{figure}

Fig.~\ref{lightcurves} shows the 14 light curves in four bands (GG420, sdss $g$, no-filter and V) spanning from 2016 to 2018, phased using the spectroscopic ephemeris (Equation 2). All show variable double-hump modulations with a full amplitude of $\sim0.3$\,mag, which is compatible with a pure ellipsoidal modulation (e.g., $0.2-0.3$\,mag \citet{boc79,mcc83}). The consistent amplitude implies that the source was quiescent for all our observations. Our observations from different filters and different nights do, however, differ in detail. For brevity, we will refer to phases 0.0, 0.25, 0.5 and 0.75 as the secondary dip, the secondary hump, the primary dip, and the primary hump, respectively.

Table 5 lists minimum and maximum times and their corresponding phases, derived using parabolic fits to the light curves with uncertainties estimated by a bootstrap method. Although most phases are close to 0.0, 0.25, 0.5 or 0.75, the largest O-C of the primary hump, observed on 2018 November 07, is 26\,min from its typical phase of 0.75. However, the following secondary hump just shows a small deviation of 6\,min from phase 0.25. Due to the variations in the orbital modulation, the O-Cs of the light minima or maxima cannot be used to determine orbital period variations, as commonly used for many eclipsing DN (e.g., Z\,Cha \citep{dai09} and V2051\,Oph \citep{qia15}).

\begin{figure}
\centering
\includegraphics[width=14.0cm]{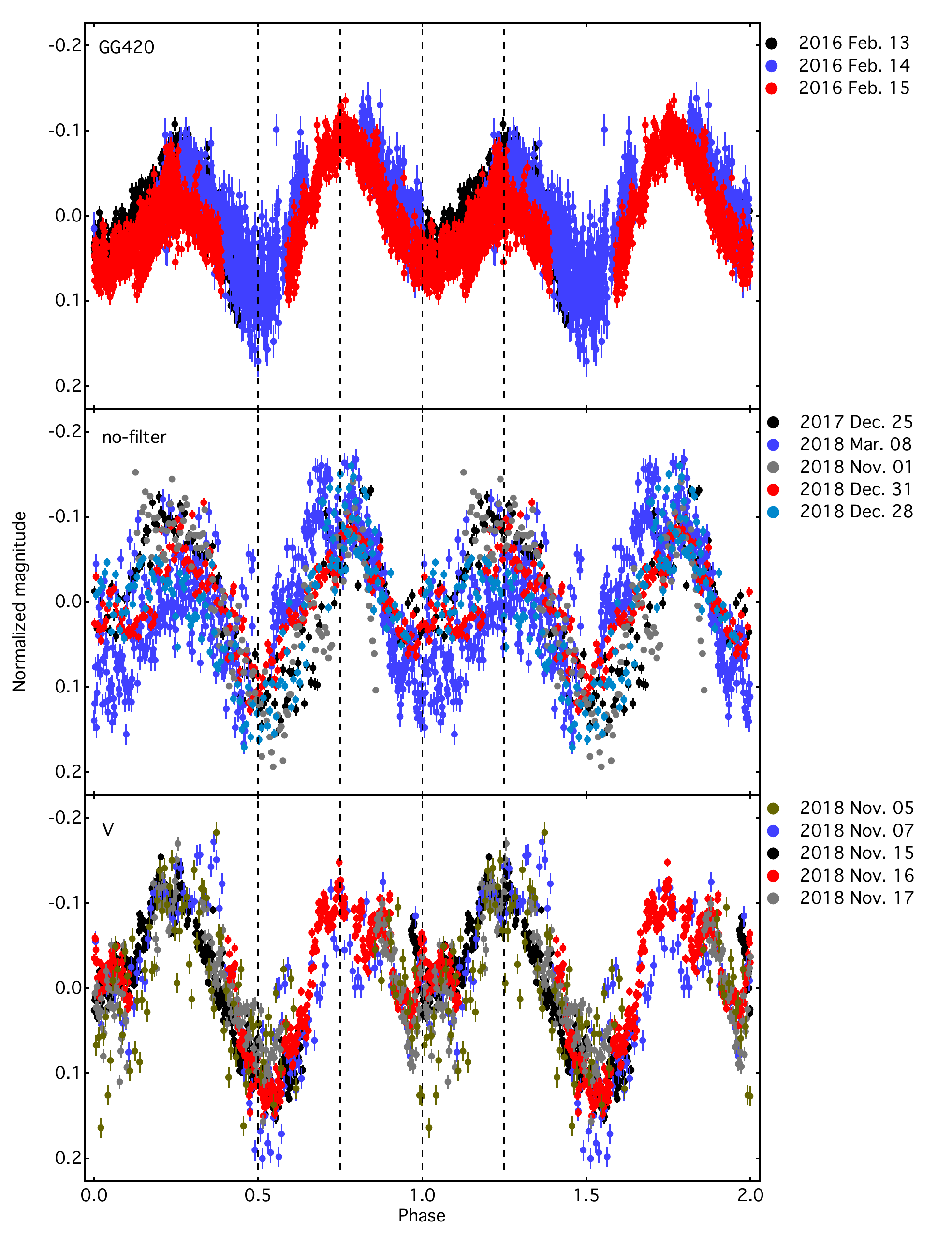}
\caption{\small{Thirteen normalized and phased light curves grouped into three bands: GG420, no-filter and V. The sdss $g$ light curve obtained on 2016 November 29 is omitted. The different nights are distinguished by the colors listed in the legends.}}
\label{filtercurves}
\end{figure}

The double-hump modulation is less distinct in the single sdss $g$-band light curve observed on 2016 November 29, due to the large scatter and relatively short duration of the observation. Fig.~\ref{filtercurves} shows the normalized and phased light curves in the other three bands superposed to demonstrate the variations in the orbital modulation in the different bands. The three sequential days of GG420-band light curves show more typical double-hump modulations with a higher-level primary hump (0.75) and lower-level primary dip (0.5), similar to other dwarf nova (e.g., J0632+2536 and TW\,Vir \citep{dai18}). However, the three no-filter-band light curves obtained 2017 December 25, 2018 November 01 and December 31 (panels c, d and h in Fig.~\ref{lightcurves}) display two humps at the same flux level and much deeper (by $\sim0.1$\,mag) primary dips. This is more like a pure ellipsoidal modulation with equal maxima and the deepest minimum around phase 0.5 caused by the tidal distortion of the Roche-lobe-filling secondary star \citep{boc79}. For KZ\,Gem, the amplitude of the V-band ellipsoidal variation is larger than that of the long-period DN TT\,Crt ($<$0.2\,mag) derived by \citet{szk92}.

Panel d of Fig.~\ref{lightcurves} shows that the lower-level secondary hump (0.25) detected on 2018 March 08 can rise to the level of the primary hump in eight months. Compared with the no-filter-band ellipsoidal modulation on 2018 December 31 (panel h of Fig.~\ref{lightcurves}), the light curve observed three days earlier (panel g of Fig.~\ref{lightcurves}) with the same filter and telescope (XLO 2.16m) clearly shows the typical double-hump modulation. Although the three V-band light curves plotted in panel f of Fig.~\ref{lightcurves} display ellipsoidal modulation similar to the no-filter-band light curves, their individual descending branches around phases $0.25-0.5$ show slight variations in the primary dip or secondary hump. The V-band light curve with a long duration of 6.22\,hr taken on 2018 November 07 presents an atypical modulation in that the primary hump is lower than the secondary hump, similar to the long-period CV CXOGBS\,J174444.7−260330 \citep{rat13}. The normalized V-band light curves superposed in the bottom panel of Fig.~\ref{filtercurves} confirm this atypical double-hump modulation, which is opposite to the modulations in GG420 and no-filter bands (the upper two panels of Fig.~\ref{filtercurves}). In summary, the orbital modulation of KZ\,Gem sometimes appears as a pure ellipsoidal modulation, and sometimes switches to a typical DN double-hump modulation. 

\subsubsection{Transient Events}

\begin{figure}
\centering
\includegraphics[width=14.0cm]{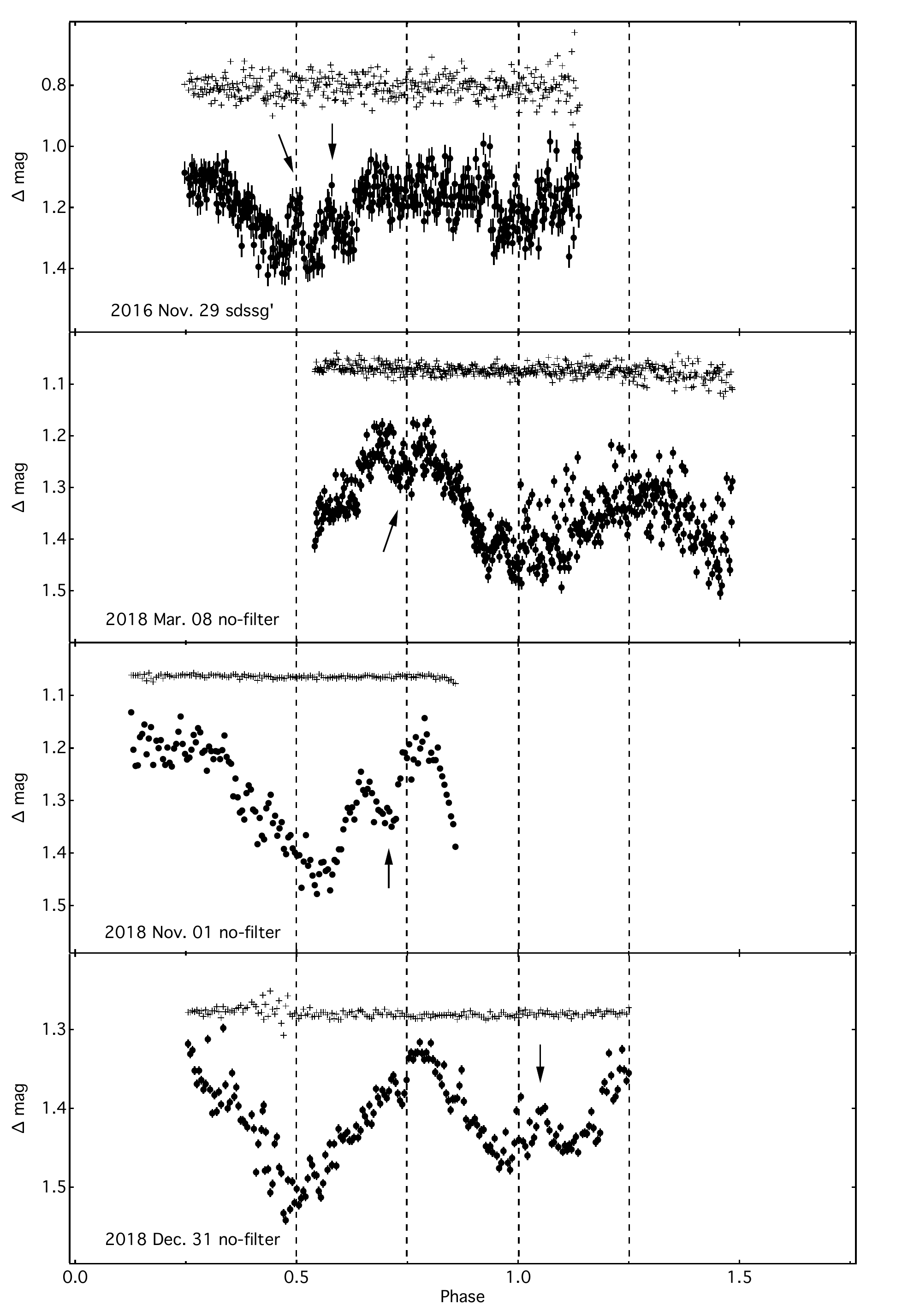}
\caption{\small{Four phased light curves showing five transient events (arrows) in time order from top to bottom.}}
\label{transientcurves}
\end{figure}

Although transient events have been detected in many low-state magnetic CVs \citep[e.g.][]{kaf09,ara05}, similar reports for quiescent dwarf nova are rare. Inspection of Fig.~\ref{transientcurves} shows 5 transient events that have an almost symmetric profile (i.e., nearly equal rise and fall) rather than an exponential profile as detected in typical X-ray/optical flares found in polars \citep[e.g.][]{dai13,ter10}. Since two of the five events are dips rather than brightening events (humps), a simple parabolic function may be more appropriate to fit them than a Gaussian function. The details of the transient events as listed in Table 6 show that the observed symmetric events last longer and have smaller amplitudes than those detected in the low-state polar AM\,Her \citep{dai13}.

The $g$-band light curve shown in the top panel of Fig.~\ref{transientcurves} displays two sequential brightening events with the nearly equal timescale of $\sim30$\,min. The peak of the first one almost occurs at the light minimum of the primary dip, while the second has a smaller amplitude. This resembles the two R-band brightening events with similar amplitudes and durations that occurred at the beginning of the low-to-high state transition in the prototype polar AM\,Her \citep{dai13}. Since most of the brightening events (flares) detected in cool stars showing typical exponential profiles are detected in a red R band and have a range of amplitude $0.02\sim0.3$\,mag \citep[e.g.][]{qia12,zha10}, the twin brightening events in the blue sdss $g$ band may not originate from the secondary star of KZ\,Gem, but are likely related to the white dwarf or the inner part of the disk. The three brightening events of KZ\,Gem have shorter durations, smaller amplitudes, and more symmetric profiles than the quasi-periodic mini-outbursts detected in Kepler observations of two other DN (V1504\,Cyg \citealt{osa14}, and CRTS\,J035905.9+175034 \citealt{lit18}), which had a duration of $\sim2$\,d, amplitudes of $\sim0.5$\,mag and irregular morphologies. This implies that the brightening events might have a different origin than the mini-outbursts of the other two DN. In the bottom panel of Fig.~\ref{transientcurves}, a brightening event in the no-filter band occurring at phase 0.04 (close to the light minimum of the secondary dip), has the longest duration of 64\,min and the smallest amplitude of 0.093\,mag.

The two significant dips in the no-filter band with similar amplitude and duration observed on 2018 March 08 and November 01 are shown in the middle two panels of Fig.~\ref{transientcurves}. The former appears at the peak of the primary hump (0.75), while the latter is slightly asymmetric and occurs $\sim20$\,min ahead of phase 0.75 (i.e., the egress branch of the V-shaped primary hump). In the no-filter-band light curve detected on 2018 December 31, a less distinct dip with a short duration ($\sim9$\,min) and small amplitude ($\sim0.05$\,mag) appears at phase 0.74. Thus, the detected brightening and dip events in KZ\,Gem at quiescence seem to be related to the secondary light minima and the primary light maxima, respectively.

\subsection{Synthetic analysis}

\citet{dai18} proposed a phenomenological model to reproduce the double-hump light curves of low-inclination DN, and successfully obtained photometric solutions for three DN by using the light curve code XRBinary developed by E. L. Robinson\footnote{\url{http://www.as.utexas.edu/~elr/Robinson/XRbinary.pdf}} and the nonlinear fitting code NMfit. This demonstrated the reliability of the model obtained using the codes. We attempted to use this model to carry out a complete synthetic analysis for KZ\,Gem based on our followup ground-based light curves spanning $\sim3$ years.

All 13 light curves with the GG420, no-filter, and V bands were separated into three types. Since the no-filter-band and V-band light curves consist of data observed by many different telescopes on several separated nights, they show larger scatter than those in the three-day sequential GG420-band light curves with a total of $\sim4000$ data points. Due to the higher orbital phase resolution and smaller scatter, we derived a light curve model for KZ\,Gem from the overlapped GG420-band light curve shown in the top panel of Fig.~\ref{filtercurves}. Before running XRBinary, the light curve was first binned with a phase resolution of 0.01. Since the main feature of the orbital light curve is the variable ellipsoidal modulation, the two stellar component stars dominate the flux contributions to the light curve, like the DN TT\,Crt \citep{szk92} and J0632+2536 \citep{dai18}. At times, one or more hotspots on the disk may change the orbital modulation from a pure ellipsoidal modulation to a typical DN double-hump modulation. The amplitude of the H$\alpha$ emission line variation in the greyscale plot (bottom panel of Fig.~\ref{mdmspecs}) is evidence of an S-wave indicating the motion of a hotspot on the disk similar to the trailed spectra of the DN SDSS\,J0116+09 \citep[see Fig.~3 of][]{szk18}. Hence, two models, model-1 (no surface hotspot) and model-2 (a surface hotspot) were investigated.

\subsubsection{System parameters for KZ\,Gem}

Based on the data from RKcat and other literatures, the averaged white dwarf masses for three grades (A: well-measured; B: less-well-measured; C: without error bars) are 0.82\,$M_{\odot}$, 0.78\,$M_{\odot}$ and 0.74\,$M_{\odot}$, respectively. They are similar to the previous prediction of 0.83$\pm$0.23\,$M_{\odot}$ for CVs in average \citep{zor11}. Because the white dwarf mass of KZ\,Gem is not accurately determined, we preset the initial $M_{\rm wd}$ to 0.83\,$M_{\odot}$ as a starting point of the iterations, rather than a fixed or assumed parameter. Since the derived orbital period of KZ\,Gem is above the period gap, an average CV white dwarf temperature of $\sim25,793$\,K derived by \citet{sio99,urb06} was assumed to be the initial $T_{\rm wd}$. The NaD absorption line shown in the two XLO and mean MDM spectra (Fig.~\ref{xlostar3} and \ref{mdmspecs}) implies a late K star in KZ\,Gem. The initial temperature and mass of the secondary star ($T_{\rm rd}$ and $M_{\rm rd}$) was assumed to be 4410\,K and 0.67\,$M_{\odot}$, respectively. Since a preparatory accretion disk model derived by fixing the masses and temperatures of the two component stars to be the initial parameters in the NMfit program implies the accretion disk almost extends to the surface of the white dwarf, $R_{\rm in}$ is always equal to $R_{\rm wd}$ during the iterations.

Based on the phased and binned light curve in GG420 band, all 17 (13 for model-1) parameters were set to be adjustable. With model-1, the best-fit light curve shows a large deviation from the observed light curve at phases $0.0-0.25$. However, this deviation can be eliminated using model-2. A much smaller $\chi^{2}$ indicates that model-2 with two default limits: $R_{\rm in}$=$R_{\rm wd}$ and 0.74\,$\leq$q$\leq$\,0.88 gives a better fit than model-1. The best-fitting parameters and their uncertainties were estimated by the code NMfit. The modeled light curve in the GG420 band is plotted in the top left-hand panel of Fig.~\ref{models}. For KZ\,Gem, the irradiation effect is not calculated by XRBinary as $L_{\rm rd}>L_{\rm wd}+L_{\rm d}$. Although an accurate white dwarf mass cannot be obtained from our light curves due to the almost constant flux contributions from the white dwarf ($\sim4\%$) to the system light, the appropriate physical parameters of the white dwarf can be roughly limited within a small range of q. A search using white-dwarf mass covering a range from 0.6-1.1\,$M_{\odot}$ indicates that the $\chi^{2}$ only shows a small change from 14 to 17, and the minimal $\chi^{2}$ is found at 0.86\,$M_{\odot}$ consistent with the average mass for CVs found from the detailed study by \citet{zor11}. Based on the MK spectral classes \citep{cox00}, a normal K0V star is estimated from $T_{\rm rd}=5120(\pm110)$\,K. The derived $M_{\rm rd}\sim0.7(\pm0.2)\,M_{\odot}$ indicates a later K type dwarf. Hence, our light curve model demonstrates that the secondary of KZ\,Gem is an early K-type dwarf in accord with the observed spectra. Like the other long period DN TW\,Vir and J0632+2536 \citep{dai18}, $M_{\rm rd}$ and $R_{\rm rd}$ of KZ\,Gem are basically consistent with the semi-empirical mass-radius relation of CV donor sequence \citep{kni06,kni11}, and the CVs mass/radius-period relations \citep{war03,smi98} shown in the bottom left-hand and two right-hand panels of Fig.~\ref{secondarygraphs}, respectively. To verify the mass of the secondary star derived in this synthetic model, further high-resolution observations are needed. The top left-hand panel of Fig.~\ref{secondarygraphs} indicates that $T_{\rm rd}$ of KZ\,Gem is $\sim900$\,Kelvin higher than the prediction of the semi-empirical CV donor sequence \citep{kni06,kni11}. This implies that the secondary star has undergone some nuclear evolution. The other three DN investigated by \citet{dai18} shown in the top left-hand panel of Fig.~\ref{secondarygraphs} may also have evolved donors similar to KZ\,Gem. Hence, these four DN appear to be Peculiar Cataclysmic Variables (PCVs) containing evolved donors \citep{reb14,ren18}.

\subsubsection{Disk models in three bands}

\begin{figure}
\centering
\includegraphics[width=16.0cm]{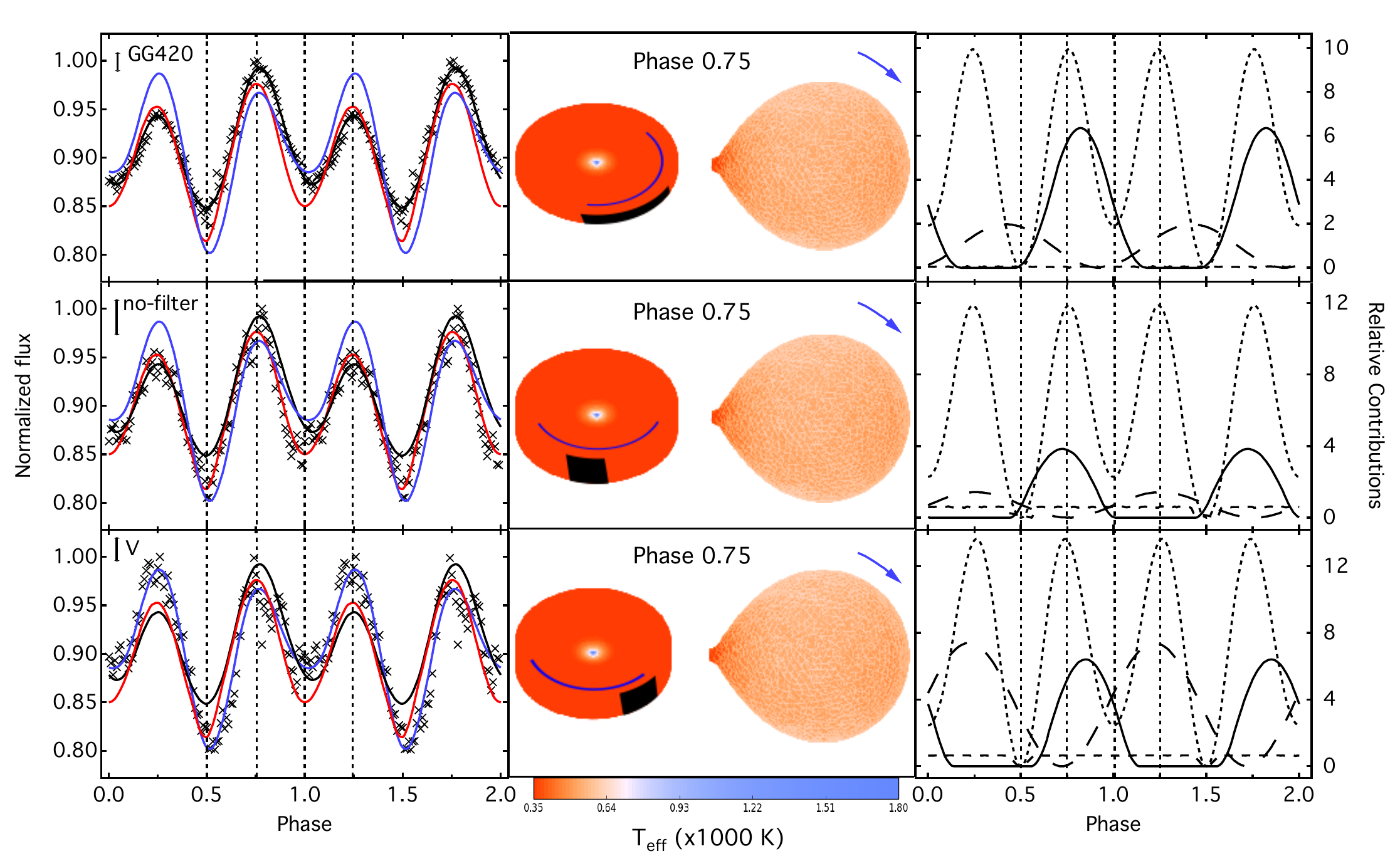}
\caption{\small{(Left panels): Phased, normalized and binned light curves in GG420, no-filter, and V, superimposed on the best-fitting light curves (black for GG420, red for no-filter and blue for V). The error bars in the top-left corners indicate the data scatter of the binned light curves. (Middle panels): The 2D binary configurations at phase 0.75 computed using Phoebe 2.0 from the fits to the different filters' data. The arrow shows the clockwise rotation of the binary. The colors in the 2D CV configuration denote the effective temperatures. The hotspot at the edge of the disk is filled in with black, rather than the color picked from the color bar, because the small temperature difference between the hotspot and the neighboring region of the disk reduces the contrast of the hotspot. (Right panels): Relative flux contributions from different components. The dotted and short dashed lines refer to the contributions from the two stellar components (white and red dwarfs) and the disk without the hotspots, respectively. The solid and long dashed lines denote the contributions from hotspot$^{\rm es}$ and hotspot$^{\rm ss}$, respectively.}}
\label{models}
\end{figure}

\begin{figure}
\centering
\includegraphics[width=14.0cm]{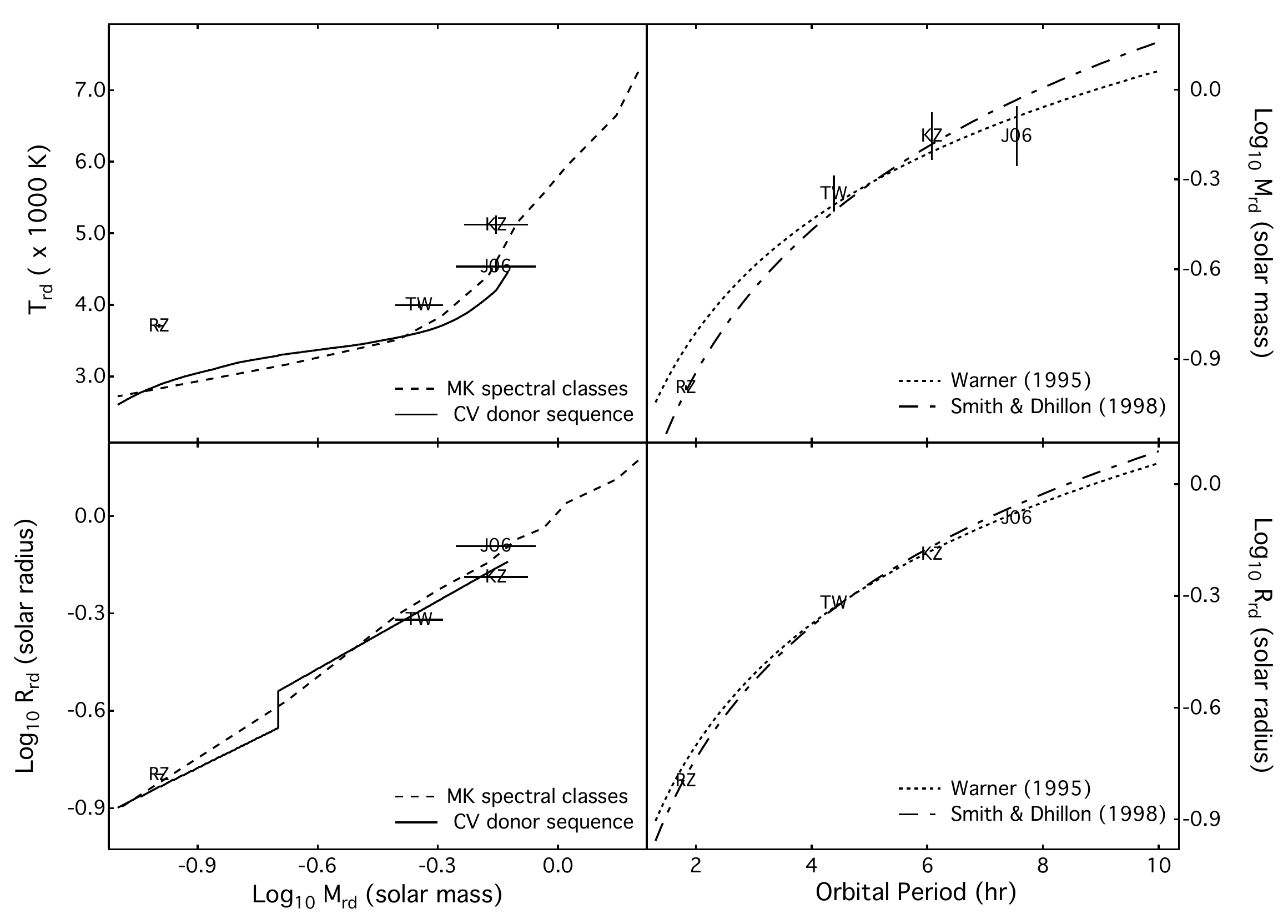}
\caption{\small{Four relationships of the secondaries. (Top left-hand panel): the relationship between the logarithm of mass and the effective temperature. (Bottom left-hand panel): the logarithm of mass-radius relationship. (Top right-hand panel): the period–mass relationship. (Bottom right-hand panel): the period–radius relationship. The dashed and solid lines denote the relationships based on the isolated low-mass stars \citep{cox00}, and the semi-empirical CV donor sequence \citep{kni06,kni11}, respectively. The dashed-dotted and dotted lines describe the relationships derived by \cite{smi98} and \cite{war03}, respectively. The data point marked by RZ, TW, J06, and KZ refer to the four DN: RZ\,Leo, TW\,Vir, J0632+2536 and KZ\,Gem, respectively.}}
\label{secondarygraphs}
\end{figure}

Five best-fitting system parameters (q, $i$, $M_{\rm wd}$, $T_{\rm wd}$ and $T_{\rm rd}$) derived from the GG420-band light curve were fixed to model the 12 parameters of the disk in three bands. The derived parameters listed in Table 7 are used to visualize a system configuration of KZ\,Gem using Phoebe 2.0\footnote{The version of Phoebe used for the CV plotting is 2.0a2.}. All three 2D CV configurations at phase 0.75 shown in the middle panels of Fig.~\ref{models} indicate a consistent disk model consisting of a large, cool and thick disk with a flat temperature distribution around a central white dwarf with a mass of 0.86($\pm$0.09)\,$M_{\odot}$, and two hotspots (one at the vertical side of the edge of the disk, which we will call hotspot$^{\rm es}$ and the other one on the surface of the disk (hotspot$^{\rm ss}$)). The luminosity of hotspot$^{\rm es}$ is given by $\dot{M}_{\rm rd}\,\simeq\,L_{\rm acc}R_{\rm out}/GM_{\rm wd}$. From this, we roughly estimate a range of mass transfer rate to be $2.7-7.9\times10^{-10}\,M_{\odot}yr^{-1}$ due to the different $L_{\rm acc}$ in different bands. This corresponds to a mass loss timescale (i.e., $\tau_{\dot{M}}\simeq\,M_{\rm rd}/\dot{M}_{\rm rd}$) in the range of $0.9-2.5\times10^{9}$\,yr, far larger than the thermal (or Kelvin-Helmholtz) timescale of the secondary star ($\tau_{\rm kh}\sim9\times10^{7}$\,yr). This is generally the case for CV secondaries \citep{pat84}. The secondary star is somewhat smaller in radius than an isolated main sequence star of the same mass, which may imply an overestimated $\dot{M}_{\rm rd}$. Therefore, the mass transfer via the L1 point should be much slower and the secondary is always able to maintain thermal equilibrium.

The right-hand panel of Fig.~\ref{models} shows the relative flux contributions in percentage from the different model components calculated by XRBinary. The size of the disk in the three bands is similar, while the thickest disk appears in the V band. Note that the $L_{\rm d}$ in the no-filter band is only $\sim30\%$ of that in the GG420 and V bands due to the cooler disk and smaller hotspot$^{\rm es}$ in the no-filter band. The zero points of the relative flux contributions from the different components listed in Table 8 indicate that two stellar components showing a pure ellipsoidal modulation dominate the system light in the no-filter band (i.e., the maximal percentage is close to 95\%). Despite this, a weak hotspot$^{\rm es}$ at phase $0.72(\pm0.03)$ distorts the pure ellipsoidal modulation into the typical double-hump modulation in the no-filter band. However, the averaged orbital modulation in the V band (with the relative flux contributions from the two component stars similar to that in the GG420 band), is an almost pure ellipsoidal modulation rather than the typical double-hump modulation. Investigation of the middle panels of Fig.~\ref{models} indicates that the hotspot$^{\rm es}$ in the GG420 band is larger than that in the V band, while the hotspot$^{\rm ss}$ in the V band with a similar size is located ahead of the one in the GG420 band. Moreover, the hotspot$^{\rm es}$ and hotspot$^{\rm ss}$ in the V band have roughly equal contributions to the primary and secondary humps. This may explain the nearly equal maxima of the two humps described by a pure ellipsoidal modulation. Hence, the geometric sizes, positions, and intensities of two hotspots may be the key parameters distorting the light curve from that of a pure ellipsoidal modulation. All three light curve models demonstrate that an ellipsoidal modulation caused by a K type dwarf dominates the orbital modulation of KZ\,Gem, and the variations in the two orbital humps result mainly from the two hotspots.

\subsubsection{Comparison with the Gaia data}

According to Equation (1) of \citet{dai18}, a V-band magnitude of KZ\,Gem estimated from three parameters: the Gaia distance $D_{\rm g}$, the system luminosity $L_{\rm all}=L_{\rm rd}+L_{\rm wd}+L_{\rm d}$ and a model-dependent bolometric correction $BC_{\rm v}$, can be compared to the result obtained from the Gaia mission \citep{gai16}. 

The Gaia parallax of KZ\,Gem implies a distance $D_{\rm g}=1293\pm149$\,pc \citep{lur18}. Since BC$_{\rm v}$ listed in three tabulations \citep{flo96,bes98,cas14} are almost the same due to $T_{\rm rd}$$>$\,4000\,K, we set BC$_{\rm v}$ to $-0.22(\pm0.03)$, interpolated from the updated BC$_{\rm v}$ table proposed by \citet{cas14}. The three disk models in different bands show different $L_{\rm d}$, but a small range of $16.70-16.87$\,mag for KZ\,Gem can be estimated from $L_{\rm all}$=1.11-1.29$\times10^{33}$\,erg\,s$^{-1}$ due to the domination of $L_{\rm rd}$=1.0$\times10^{33}$\,erg\,s$^{-1}$ ($>$75\% of the system light). Although the apparent visual magnitude of KZ\,Gem (S3) is not listed in SIMBAD, the estimated V-band magnitude is close to B=16.8\,mag listed in RKcat, sdss $g$=16.74\,mag and sdss $r$=16.43\,mag listed in SDSS. Thus, the derived synthetic model of KZ\,Gem is compatible with the Gaia distance.

Like the two DN J0632+2536 and TW\,Vir \citep{dai18}, the $T_{\rm eff}$ of KZ\,Gem shown in the Gaia catalog is higher than the derived $T_{\rm rd}$=5120($\pm$110)\,K listed in Table 7. This further confirms the previous speculation by \citet{dai18} that a higher $T_{\rm eff}$ derived by Gaia is common for DN due to possible contributions from hotter components in these systems (e.g., the white dwarf and disk). Note that the Gaia temperatures are determined from three broad bandpasses \citep{and18} and the DR2 release notes urge caution in using them \footnote{\url{https://gea.esac.esa.int/archive/documentation/GDR2/pdf/GaiaDR2_documentation_1.0.pdf}}.

\section{Conclusions}

Cross-checking several CV databases revealed a coordinate discrepancy of $9''$ for the DN KZ\,Gem. We obtained spectra of four nearby targets which show that the correct identification is a star matching the coordinates listed by RKcat and VSX.

We used absorption-line radial velocities collected from 2014 to 2019 to improve the orbital period to $0.2224628(2)$\,d. We collected light curves that show a variable double-hump modulation with a typical amplitude of $\sim0.3$\,mag. This orbital modulation shows night-to-night variations. We also detected three brightening and two dipping transient events with nearly symmetric rises and declines appearing around the orbital minima and maxima. The phased light curves are consistent with the high-precision spectral ephemeris.

The binned and normalized light curve in the GG420 band was analyzed using the light curve code XRBinary and nonlinear fitting code NMfit. The best-fitting synthetic model indicates that KZ\,Gem is composed of a primary white dwarf (0.86($\pm$0.16)\,$M_{\odot}$) and a Roche-lobe-filling K type dwarf (0.7($\pm$0.2)\,$M_{\odot}$) with an higher effective temperature (5120($\pm$110)\,K) than the typical CV secondaries at the same orbital period \citep{kni06,kni11}, orbiting each other with an orbital inclination of 51$^{\circ}$.6($\pm$1$^{\circ}$.4). The hotter secondary star implies that KZ\,Gem may be a new candidate of PCVs. By fixing the system parameters, a consistent disk model with a large, cool and thick disk with a flat temperature distribution including two hotspots near phase 0.75 was achieved for three bands of light curves. We estimate $\dot{M}_{\rm rd}$ to be 2.7-7.9$\times$$10^{-10}$\,$M_{\odot}$$yr^{-1}$. All derived light curve models are in accord with results calculated from Gaia DR2 and indicate that a pure ellipsoidal modulation caused by a K type dwarf dominates the orbital modulation.

\acknowledgments

This work was partly supported by CAS Light of West China Program, the Chinese Natural Science Foundation (Nos. 11133007 and 11325315), and the Science Foundation of Yunnan Province (No. 2016FB007). PS acknowledges support from NSF grant AST-1514737. We acknowledge the support of the staff of the Xinglong 2.16m and 0.85m telescopes. This work was partially supported by the Open Project Program of the Key Laboratory of Optical Astronomy, National Astronomical Observatories, Chinese Academy of Sciences. Based on observations obtained with Apache Point Observatory 3.5m and the 0.5m Astrophysical Research Consortium Small Aperture Telescope, and the 1.3m and 2.4m at the MDM Observatory, operated by Dartmouth College, Columbia University, Ohio State University, Ohio University, and the University of Michigan. We thank Wang Huijuan (王汇娟) and Ren Juanjuan (任娟娟), for their great assistance on obtaining and reducing two XLO spectra taken on 2018 January 09 and 10.

\software{IRAF \citep{tod86,tod93}, XRBinary \citep[v2.4;][]{dai18}, NMfit \citep[v2.0;][]{dai18}, Phoebe \citep[v2.0;][]{prs16})}

\begin{table}
\caption{Coordinates and magnitudes of the 6 stars nearby KZ\,Gem listed in SDSS DR12$^{\tablenotemark{a}}$}
\begin{center}
\begin{tabular}{lccccccc}
\hline\hline
Name & R. A. & decl. & $u$ & $g$ & $r$ & $i$ & $z$\\
\hline
S1$^{\tablenotemark{b}}$ & 06:53:02.74 & +16:39:49.87 & 19.94(4) & 18.67(1) & 18.15(1) & 17.96(1) & 17.90(2)\\
S2 & 06:53:02.70 & +16:39:53.98 & 21.66(12) & 18.96(1) & 17.56(1) & 16.93(1) & 16.60(1)\\
S3$^{\tablenotemark{c}}$ & 06:53:02.45 & +16:39:58.60 & 17.54(1) & 16.74(-) & 16.43(-) & 16.30(1) & 16.27(1)\\
S4 & 06:53:02.50 & +16:40:05.99 & 16.22(1) & 14.95(-) & 14.59(-) & 14.48(-) & 14.49(-)\\
S5 & 06:53:01.81 & +16:39:57.45 & 20.74(6) & 19.38(1) & 18.80(1) & 18.55(1) & 18.53(3)\\
S6 & 06:53:03.24 & +16:39:56.15 & 21.50(10) & 19.32(1) & 18.23(1)	& 17.84(1) & 17.66(2)\\
\hline\hline
\end{tabular}
\end{center}
\tablenotetext{a}{\,\cite{ala15}.}
\tablenotetext{b}{\,Coordinates specified by the SDSS and SIMBAD.}\tablenotetext{c}{\,Coordinates listed in RKcat and VSX.}
\end{table}

\begin{table}
\caption{Summary of observations}
\begin{center}
\begin{tabular}{llcc}
\hline\hline
UT Date & Observations & Filters/Spectrographs$^{\tablenotemark{a}}$ & Exposures\\
\hline
2014 Dec. 14 & MDM 2.4m & modspec & $2\times600s$ spectra\\
2016 Feb. 13-15 & MDM 1.3m & GG420 & $930\times10s$ images\\
2016 Nov. 29 & APO 0.5m & sdss $g$ & $415\times30s$ images\\
2016 Jan. 14-20 & MDM 2.4m & modspec & $25\times600s$ spectra\\
2016 Feb. 11-15 & MDM 2.4m & modspec & $5\times600s$ spectra\\
2016 Dec. 05$^{\tablenotemark{b}}$ & APO 3.5m & DIS & $1\times600s$ spectrum\\
2017 Jan. 22$^{\tablenotemark{c}}$ & XLO 2.16m & BFOSC & $1\times3600s$ spectrum\\
2017 Dec. 25 & XLO 0.85m & no-filter & $149\times120s$ images\\
2018 Jan. 09 & XLO 2.16m & BFOSC & $1\times2400s$ spectrum\\
2018 Jan. 10 & XLO 2.16m & BFOSC & $1\times3600s$ spectrum\\
2018 Jan. 15$^{\tablenotemark{b}}$ & APO 3.5m & DIS & $2\times900s$ spectra\\
2018 Jan. 17 & APO 3.5m & DIS & $6\times600s$ spectra\\
2018 Jan. 19 & APO 3.5m & DIS & $3\times600s\,+\,9\times900s$ spectra\\
2018 Mar. 08 & XLO 0.85m & no-filter & $502\times30s$ images\\
2018 Nov. 01 & XLO 0.85m & no-filter & $148\times90s$ images\\
2018 Nov. 05 & APO 0.5m & V & $101\times150s$ images\\
2018 Nov. 07 & APO 0.5m & V & $137\times150s$ images\\
2018 Nov. 14-20 & MDM 2.4m & modspec & $6\times1200s$ spectra\\
2018 Nov. 15-17 & MDM 1.3m & V & $420\times30s$ images\\
2018 Dec. 13-15 & MDM 2.4m & OSMOS & $10\times1200s$ spectra\\
2018 Dec. 28 & XLO 2.16m & no-filter & $165\times120s$ images\\
2018 Dec. 31 & XLO 2.16m & no-filter & $201\times90s$ images\\
2019 Jan. 14 & MDM 2.4m & OSMOS & $3\times900s$ spectra\\
2019 Jan. 16 & MDM 2.4m & OSMOS & $4\times600s$ spectra\\
\hline\hline
\end{tabular}
\end{center}
\tablenotetext{a}{\,Two spectra simultaneously taken by the DIS with the blue and red channels covering two wavelength ranges of 3900-5040\,\AA and 6225-7400\,\AA, respectively. The wavelength ranges of spectra taken by the OSMOS and BFOSC are 4200-7500\,\AA and 4600-8400\,\AA, respectively. A GG420 filter blocks light at wavelengths less than 4200\,\AA.}
\tablenotetext{b}{\,Spectra for S1.}
\tablenotetext{c}{\,Simultaneous spectra for S1, S2 and S4.}
\end{table}

\begin{table}
\caption{S3 emission line equivalent widths}
\begin{center}
\begin{tabular}{ccc}
\hline\hline
UT Date & H$\alpha$ (\AA) & H$\beta$ (\AA)\\
\hline
2018 Jan 09 & 7.63(1) & --\\
2018 Jan 10 & 6.28(2) & --\\
2018 Jan 17 & 8.53(1) & 1.67(1)\\
& 8.89(1) & 3.13(1)\\
& 10.47(2) & 3.52(1)\\
& 10.21(2) & 2.78(1)\\
& 10.77(1) & 3.90(1)\\
& 8.68(2) & 3.06(1)\\
2018 Jan19 & 7.52(4) & --\\
& 5.17(2) & --\\
& 5.23(3) & --\\
& 5.62(8) & --\\
& 5.18(8) & --\\
& 4.53(1) & --\\
& 6.49(2) & --\\
& 7.31(3) & --\\
& 6.20(4) & --\\
& 5.44(7) & --\\
& 6.49(3) & --\\
& 5.26(3) & --\\
\hline\hline
\end{tabular}
\end{center}
\end{table}

\begin{table}
\caption{Parameters of the sine fits for the radial velocities of KZ\,Gem}
\begin{center}
\begin{tabular}{ccccc}
\hline\hline
& $\gamma$ & K & $T_{\rm 0}$ & rms\\
& km\,s$^{-1}$ & km\,s$^{-1}$ & BJD (2450000+) & km\,s$^{-1}$\\
\hline
Absorption lines$^{\tablenotemark{a}}$ & $-13(\pm3)$ & $191(\pm5)$ & 7434.9280(7) & 14\\
H$\alpha$ emission lines$^{\tablenotemark{b}}$ & $-15.7(\pm1.6)$ & $154(\pm3)$ & 8137.683(5) & 44\\
\hline\hline
\end{tabular}
\end{center}
\tablenotetext{a}{\,55 MDM spectra spanning from 2014 December to 2019 January.}
\tablenotetext{b}{\,18 APO spectra taken on 2018 January 17 and 19.}
\end{table}

\begin{table}
\caption{Light minimum and maximum times of KZ\,Gem}
\begin{center}
\begin{tabular}{llccc}
\hline\hline
UT Date & BJD & Phase & hump/dip$^{\tablenotemark{a}}$ & O-C\\
\hline
& 2450000+ &&& min\\
\hline
2016 Feb. 13 & 7431.8233(4) & 0.044(4) & Sec.d & 14\\
& 7431.8700(3) & 0.254(3) & Sec.h & 1\\
2016 Feb. 15 & 7433.6524(3) & 0.266(3) & Sec.h & 5\\
& 7433.7648(2) & 0.771(3) & Pri.h & 7\\
& 7433.8253(2) & 0.043(3) & Sec.d & 14\\
2016 Nov. 29 & 7722.905(1) & 0.495(6) & Pri.d & -2\\
2017 Dec. 25 & 8103.2044(9) & 0.992(5) & Sec.d & -3\\
& 8103.2591(5) & 0.238(4) & Sec.h & -4\\
& 8103.328(1) & 0.547(6) & Pri.d & 15\\
2018 Mar. 08 & 8186.0150(4) & 0.236(4) & Sec.h & -4\\
& 8186.0834(8) & 0.544(5) & Pri.d & 14\\
& 8186.135(2) & 0.78(1) & Pri.h & 10\\
2018 Nov. 01 & 8424.3405(8) & 0.541(5) & Pri.d & 13\\
2018 Nov. 05 & 8428.899(2) & 0.03(1) & Sec.d & 10\\
& 8429.005(4) & 0.51(2) & Pri.d & 3\\
2018 Nov. 07 & 8430.856(3) & 0.83(1) & Pri.h & 26\\
& 8430.954(1) & 0.270(6) & Sec.h & 6\\
2018 Nov. 15 & 8438.9545(3) & 0.233(3) & Sec.h & -5\\
& 8439.0259(7) & 0.554(4) & Pri.d & 17\\
2018 Nov. 16 & 8439.9119(3) & 0.537(3) & Pri.d & 12\\
& 8439.9683(4) & 0.790(4) & Pri.h & 13\\
2018 Dec. 28 & 8481.281(1) & 0.496(5) & Pri.d &-1\\
2018 Dec. 31 & 8484.177(2) & 0.51(1) & Pri.d & 3\\
& 8484.2352(4) & 0.776(4) & Pri.h & 8\\
& 8484.291(1) & 0.027(6) & Sec.d & 9\\
& 8484.3466(9) & 0.277(5) & Sec.h & 9\\
\hline\hline
\end{tabular}
\end{center}
\tablenotetext{a}{\,Pri.h, Pri.d, Sec.h and Sec.d refer to the primary hump, primary dip, secondary hump and dip around phases 0.75, 0.5, 0.25 and 0.0, respectively.}
\end{table}

\begin{table}
\caption{Transient excursions of KZ\,Gem}
\begin{center}
\begin{tabular}{ccccc}
\hline\hline
UT Date & Filter & BJD (Phase)$^{\tablenotemark{a}}$ & Duration & Amplitude$^{\tablenotemark{b}}$\\
&& 2450000+ & min & mag\\
\hline
2016 Nov. 29 & sdss $g$ & 7722.904 (0.49) & 28 & +0.31\\
&& 7722.923 (0.58) & 29 & +0.26$^{\tablenotemark{c}}$\\
2018 Mar. 08 & no-filter & 8186.018 (0.26) & 44 & -0.14\\
2018 Nov. 01 & no-filter & 8424.374 (0.69) & 49 & -0.16$^{\tablenotemark{c}}$\\
2018 Dec. 31 & no-filter & 8484.295 (0.04) & 64 & +0.093\\
\hline\hline
\end{tabular}
\end{center}
\tablenotetext{a}{\,The times (phases) of the central peaks or dips of the transients.}
\tablenotetext{b}{\,The magnitude difference between the light maximum and minimum. The minus and plus before the digits denote that the transient is a dip and hump, respectively.}
\tablenotetext{c}{\,Calculated after removing slope changes.}
\end{table}

\begin{table}
\caption{Photometric solutions for KZ\,Gem in three filters.}
\begin{center}
\begin{tabular}{lccc}
\hline\hline
Parameter\tablenotemark{a} & GG420 & no-filter & V\\
\hline
q & $0.81(\pm0.2\tablenotemark{b})$ & -- & --\\
$i$ & $51^{\circ}.6(\pm1^{\circ}.4)$ & -- & --\\
\hline
\textbf{White dwarf} &&&\\
\hline
$M_{\rm wd} (M_{\odot})$ & 0.86($\pm$0.16) & -- & --\\
$R_{\rm wd} (R_{\odot})$ & 0.0097\tablenotemark{c} & -- & --\\
$T_{\rm wd}$ & 16.9\tablenotemark{d} & -- & --\\
$L_{\rm wd}$ & 0.26\tablenotemark{c} & -- & --\\
\hline
\textbf{Red dwarf} &&&\\
\hline
$M_{\rm rd} (M_{\odot})$ & $0.7\tablenotemark{c}(\pm0.2)$ & -- & --\\
$R_{\rm rd} (R_{\odot})$ & 0.65\tablenotemark{c} & -- & --\\
$T_{\rm rd}$ & 5.12($\pm0.11$) & -- & --\\
$L_{\rm rd}$ & 10.0\tablenotemark{c} & -- & --\\
\hline
\textbf{Accretion disk} &&&\\
\hline
$R_{\rm in}^{\tablenotemark{e}} (R_{\odot})$ & 0.0097 & 0.0097 & 0.0097\\
$R_{\rm out} (R_{\odot})$ & 0.647($\pm0.004$) & 0.647($\pm0.004$) & 0.61($\pm0.03$)\\
$H_{\rm edge} (R_{\odot})$ & 0.10($\pm0.01$) & 0.238($\pm0.035$) & 0.24($\pm0.04$)\\
$\xi$ & $-0.0045^{\tablenotemark{d}}$ & $-0.0027^{\tablenotemark{d}}$ & $-0.0095^{\tablenotemark{d}}$\\
$L_{\rm d0}$ & 1.7($\pm0.2$) & 0.36($\pm0.18$) & 0.96($\pm0.18$)\\
$L_{\rm d}^{\tablenotemark{c}}$ & 2.6 & 0.8 & 2.3\\
\hline
\quad Hotspot at the edge of the disk &&&\\
\hline
$T_{\rm es}$ & 5.01($\pm0.07$) & 4.50($\pm0.22$) &4.97($\pm0.16$)\\
$\zeta_{\rm esmid}$ (phase) & 0.821($\pm0.007$) & 0.724($\pm0.03$) & 0.85($\pm0.02$)\\
$\zeta_{\rm eswidth}$ (phase) & 0.21($\pm0.02$) & 0.08($\pm0.02$) & 0.09($\pm0.02$)\\
\hline
\quad Hotspot on the surface of disk &&&\\
\hline
$\zeta_{\rm ssmin}$ (phase) & 0.73($\pm0.05$) & $0.57^{\tablenotemark{d}}$ & 0.56($\pm0.04$)\\
$\zeta_{\rm ssmax}$ (phase) & 1.11($\pm0.07$) & 0.94($\pm0.16$) & 0.88($\pm0.04$)\\
$R_{\rm ssmin} (R_{\odot})$ & 0.52($\pm0.01$) & 0.52($\pm0.01$) & 0.52($\pm0.01$)\\
$R_{\rm ssmax} (R_{\odot})$ & 0.54($\pm0.01$) & 0.54($\pm0.01$) & 0.55($\pm0.01$)\\
$T_{\rm ratio}^{ss}$ & 1.9($\pm0.1$) & 2.3($\pm0.4$) & 2.5($\pm0.1$)\\
\hline
\textbf{$\chi^{2}$} & 14.2 & 16.6 & 80.2\\
\hline\hline
\end{tabular}
\end{center}
\tablenotetext{a}{\,The unit of temperature and luminosity is $10^{3}$\,K and $10^{32}\,{\rm erg}\,{\rm s}^{-1}$, respectively.}
\tablenotetext{b}{\,The uncertainty clearly larger than the preset range of q ($0.74\sim0.88$) was obtained after removing this limit during the iterations.}
\tablenotetext{c}{\,Calculated by XRBinary.}
\tablenotetext{d}{\,Insensitive to the observed light curves.}
\tablenotetext{e}{\,Fixed in NMfit program.}
\end{table}

\begin{table}
\caption{Zero points of the relative flux contributions from different components}
\begin{center}
\begin{tabular}{lcccc}
\hline\hline
Components$\tablenotemark{a}$ & GG420 & no-filter & V\\
\hline
Binary\tablenotemark{b} & 69.3 & 82.7 & 70.4\\
Disk\tablenotemark{c} & 9.6 & 0.01 & 3.5\\
Hotspot$^{\rm es}$ & 0.0 & 0.0 & 0.0\\
Hotspot$^{\rm ss}$ & 4.8 & 1.1 & 4.5\\
\hline\hline
\end{tabular}
\end{center}
\tablenotetext{a}{\,The relative flux contributions are in percentage.}
\tablenotetext{b}{\,Only consist of white dwarf and red dwarf.}
\tablenotetext{c}{\,Accretion disk without hotspot.}
\end{table}

\clearpage

\end{CJK*}
\end{document}